\documentclass[journal,final,doublecolumn,10pt,twoside]{IEEEtranTCOM} 
\normalsize

\usepackage{flushend}
\usepackage{graphicx}
\usepackage{epstopdf}
\usepackage{subfigure}
\usepackage{enumerate}
\usepackage{url}
\usepackage{times}
\usepackage{mathtools}
\usepackage{amssymb}

\newcommand{\subparagraph}{}
\usepackage[explicit,indentafter]{titlesec}

\newcommand*{\TitleFont}{%
      \usefont{\encodingdefault}{\rmdefault}{}{n}%
      \fontsize{23}{28}%
      \selectfont}

\begin{document}

\title{\TitleFont Technical Report: Efficient Buffering and Scheduling for a Single-Chip Crosspoint-Queued Switch}

\author{Zizhong~Cao,~\IEEEmembership{Student Member,~IEEE,}
        and~Shivendra~S.~Panwar,~\IEEEmembership{Fellow,~IEEE}
\thanks{Z. Cao and S. S. Panwar are with the Department
of Electrical and Computer Engineering, Polytechnic School of Engineering, New York University, Brooklyn,
NY, 11201 USA e-mail: zc347@nyu.edu, panwar@catt.poly.edu.}}

%


\maketitle

\begin{abstract}
The single-chip crosspoint-queued (CQ) switch is a compact switching architecture that has all its buffers placed at the crosspoints of input and output lines. Scheduling is also performed inside the switching core, and does not rely on latency-limited communications with input or output line-cards. Compared with other legacy switching architectures, the CQ switch has the advantages of high throughput, minimal delay, low scheduling complexity, and no speedup requirement. However, the crosspoint buffers are small and segregated, thus how to efficiently use the buffers and avoid packet drops remains a major problem that needs to be addressed. In this paper, we consider load balancing, deflection routing, and buffer pooling for efficient buffer sharing in the CQ switch. We also design scheduling algorithms to maintain the correct packet order even while employing multi-path switching and resolve contentions caused by multiplexing. All these techniques require modest hardware modifications and memory speedup in the switching core, but can greatly boost the buffer utilizations by up to $10$ times and reduce the packet drop rates by one to three orders of magnitude. Extensive simulations and analyses have been done to demonstrate the advantages of the proposed buffering and scheduling techniques in various aspects. By pushing the on-chip memory to the limit of current ASIC technology, we show that a cell drop rate of $10^{-8}$, which is low enough for practical uses, can be achieved under real Internet traffic traces corresponding to a load of $0.9$.
\end{abstract}


%
\IEEEpeerreviewmaketitle

\section{Introduction}
\label{sec:intro}

\IEEEPARstart{I}{n} the past decade, modern Internet-based services such as social networking and video streaming have brought about a continuous, exponential growth in Internet traffic. The boom in smartphones, tablets and other portable electronic devices has made all these remote services more accessible to people, while imposing ever larger traffic burdens on the backbone networks. To accomodate the increasing demands, the capability of Internet core switches must grow commensurately. More recently, there has also been a trend to move almost everything into the cloud, and the emergence of huge data centers have brought about more challenges in data switching. Consequently, there has been continuous interest in designing high-performance switching architectures and scheduling algorithms, most of which are considered in synchronized, time-slotted systems due to high performance and ease of implementation.

Many types of switching architectures have been proposed. One of them is the output-queued (OQ) switch \cite{oq}, in which an arriving packet is always directly sent to its destination output, and then buffered there if necessary. The OQ switch may achieve $100\%$ throughput, but requires an impractically high speedup. Specifically, the switching fabric of an $N\times N$ OQ switch may need to run $N$ times as fast as the single line rate in the worst case.

Another popular kind of architecture is the input-queued (IQ) switch. In an IQ switch, packets are buffered at the input and served in a first-in-first-out (FIFO) manner. IQ switches require no speedup, but suffer from the head-of-line (HOL) blocking problem, which limits the throughput to $58.6\%$ \cite{oq}. This problem was later solved by implementing virtual output queues (VOQ) at each input. Various scheduling algorithms such as iSLIP \cite{islip} and Maximum Weight Matching (MWM) \cite{mwm01} have been proposed to achieve high throughput. However, many of these algorithms are complex, or require nearly instantaneous communications among input and output schedulers that are usually placed far apart on different line-cards due to limited on-chip memory. This can become a bottleneck for high-speed switches, in which the round-trip latency between different line-cards may span several time slots and thus is no longer negligible. For instance, the round-trip latency can be as high as about $100ns$ assuming $10m$ inter-rack cables, while each time slot lasts at most about $50ns$, assuming OC-192 or higher line speeds and $64byte$ fragmentation. A combination of IQ and OQ switches, i.e., combined-input-and-output-queued (CIOQ) switch, has also been proposed to achieve high throughput with low delay \cite{cioq}, but suffers from similar problems.

Recently, a new kind of structure called the buffered crossbar has attracted attention. Typically, one or a few buffers can be placed at each crosspoint, while others are still placed at the inputs of a switch, which effectively becomes a combined-input-and-crosspoint-queued (CICQ) switch \cite{cicq}. With the help of crosspoint buffers, scheduling becomes much easier for CICQ switches since input scheduling and output scheduling can now be performed separately. Many scheduling algorithms that support 100\% throughput and/or guaranteed service rates for IQ switches can be directly applied to CICQ switches at a lower complexity, e.g., distributed MWM algorithm DISQUO \cite{disquo}, push-in-first-out (PIFO) policy \cite{chuang05}, and smooth scheduling \cite{smoothscheduling}. On the other hand, a CICQ switch suffers from the same problem as an IQ switch due to the need for fast communications between the input line cards and the switching core. 

To avoid such implementation difficulties, Kanizo et al. \cite{cqs} consider a self-sufficient single-chip crosspoint-queued (CQ) switch whose buffering and scheduling are performed solely inside the switching core, and argue for its feasibility \cite{asic01,asic02,asic03}. According to the latest numbers, the total amount of buffer space on a single chip can be as high as $455 Mbyte$, assuming an aggressive $70\%$ memory area on a $260 mm^2$ MPU chip and a SRAM size of $0.05 \mu m^2$. Thus for a $128\times 128$ switch, each crosspoint may hold up to $455$ packets of size $64 byte$ each. However, in comparison to an IQ or OQ switch that may spread its buffer space on multiple input/output line-cards, the total buffer space of a single-chip CQ switch is still limited.

This may seem like a severe deficiency at first glance, since it has long been believed that Internet routers must provide one round-trip-time's equivalent of buffering to prevent link starvation. However, recent studies on high-speed Internet routers by Wischik and McKeown et al. \cite{sizing02,sizing04} challenge this commonly held assumption, and suggest that the optimal buffer size can be much smaller than that was previously believed. The reason lies in the fact that the Internet backbone links are usually driven by a large number of different flows, and multiplexing gains can be obtained. They also argue that short-term Internet traffic approximates the Poisson process, while long-range dependence (LRD) holds only over large time-scales. As a result, a much smaller amount of buffering is required as long as the traffic load is moderate, and thus can readily be accomodated on a single chip.

The single-chip CQ switch has many distinct features. On the one hand, using small segregated on-chip buffers instead of large aggregated off-chip memory allows much faster memory access on ASICs, which could have been a bottleneck for high speed switches. It also divides and spatially distributes the scheduling and buffering tasks to a large number of crosspoints with a low hardware requirement at each node. On the other hand, because its buffers are small and segregated, a basic CQ switch with simple scheduling algorithms, such as round-robin (RR), oldest-cell-first (OCF) and longest-queue-first (LQF), may experience far more packet drops than an IQ or OQ switch with the same total amount of buffering. Previous analyses and simulations done by Kanizo et al. \cite{cqs} and Radonjic et al. \cite{radonjic01,radonjic02} have shown that LQF provides the highest throughput for a CQ switch in many cases, but its performance is still worse than an OQ switch with the same total buffer space. This problem is more severe when there are more ports and thus the buffer size at each crosspoint is more restricted.

A key observation here is that when a certain crosspoint experiences packet overflow, other crosspoint buffers can still be quite empty, i.e., the buffer utilizations are unbalanced. The unbalanced-utilization problem becomes worse when the incoming traffic is bursty or non-uniform. As reported in \cite{cqs}, even LQF scheduling works poorly under these conditions. Unfortunately, analyses of real Internet traffic traces often reveal such burstiness and non-uniformity. As a result, how to efficiently use the crosspoint buffers so as to reduce packet drops remains a major issue before single-chip CQ switches can be widely accepted.

One possible method to lessen the problem is to add an extra \emph{load-balancing} stage in front of the original switching fabric \cite{lb02}. As incoming traffic passes through the first load-balancing stage, its burstiness and non-uniformity can be greatly reduced. However, the extra load-balancing stage can also introduce mis-sequencing, i.e., packets of the same flow may not leave in the same order as they arrive. Mis-sequencing may cause unwanted performance degradation in many Internet services and applications, e.g., TCP-based data transmission. TCP remains the most dominant transport layer protocol used in the public Internet, but it performs poorly if the correct packet order is not maintained end-to-end, because such out-of-order packets are treated as lost and trigger unnecessary retransmissions and congestion control \cite{tcp01}. As a result, many network operators insist that packet ordering must be preserved in switch design. Previous approaches to restore packet ordering include extra re-sequencing buffers \cite{lb02} and frame-based scheduling \cite{chuang05,frame02}, but at the cost of higher delay and buffer requirements.

Another candidate is \emph{deflection routing}. This concept was proposed in the networking area as early as in the 1980s. The general idea is to reroute a packet to another node or path when there is no buffer available on its regular (shortest) path. Several topologies are proposed for deflection routing, such as the Manhattan Street Network \cite{msn}. All these designs effectively share distributed buffers at different nodes and lower the packet drop rate, but they also alter the packet order due to multi-path routing. 

A third solution is \emph{buffer pooling}. Given that the crosspoint buffers are too segregated to be used efficiently, it is quite natural to consider sharing them to some extent while still preserving the flexibility of routing and ease of scheduling. Buffer sharing has been widely studied in ATM networks \cite{sm}, and been considered as a promising way to alleviate memory shortage. However, shared memory suffers from a high speedup requirement. Fairness problems may also arise, and result in a lower throughput and a higher delay \cite{partitionsurvey}.


In this paper, we investigate the effectiveness of these different approaches, and design novel switching architectures and scheduling algorithms to accomodate them onto the CQ switch. We have made some modifications to the basic CQ switch, but to what we believe to be an implementationally modest and feasible extent. 

The main contributions of this paper are as follows:

\begin{enumerate}
\item We show that the prevalent LQF policy can be inefficient in balancing the limited buffer space of CQ switches, and thus result in high packet drop rates. Three different buffer sharing techniques to improve the performances are proposed and theoretically analyzed. (Section II)

\item We propose a novel chained crosspoint-queued (CCQ) switching architecture that is suitable for load balancing and deflection routing, and jointly design buffer sharing and in-order scheduling to meet the goals of low packet drop rate and correct packet ordering. (Section III)

\item A class of pooled crosspoint-queued (PCQ) switching architecture is also investigated. We compare the sharing efficiency versus system complexity of various pooling patterns, and present effective resolution mechanisms when input/output contentions take place. (Section IV)

\item We summarize and compare all the benefits and requirements of the proposed buffer sharing techniques, and put forward a comprehensive buffer sharing solution to CQ switches under various conditions. (Section V)

\item We then extend our scope to the delay performance, support for multicast, and Quality of Service (QoS) concerns. Their applicability and implementation concerns in various CQ switches are discussed. (Section VI)

\item Extensive simulations are performed to demonstrate the effectiveness of the proposed buffering techniques and the impact of various parameters. (Section VII)
\end{enumerate}

The architecture and scheduling design of the CCQ switch is partly based on our preliminary work \cite{ancs}. However, it is not until this paper that we provide the motivation and rationale of our design, and shed more light on how the proposed buffer sharing techniques could significantly improve the performance.

In the rest of this paper, we focus on the following five switch configurations:
\begin{itemize}
\item \emph{CQ-LQF:} a basic single-stage CQ switch (Section \ref{sec:basic}) with LQF scheduling and no speedup;

\item \emph{CCQ-OCF:} a two-stage CCQ switch (Section \ref{sec:modified}) with OCF scheduling and a speedup of $2$ (Section \ref{sec:OCF});

\item \emph{CCQ-RR:} a counter-based scheme with RR scheduling that mimics \emph{CCQ-OCF} (Section \ref{sec:RR});

\item \emph{PCQ-GLQF:} a PCQ switch running generalized LQF with contention resolution at small speedups (Section \ref{sec:pooling});

\item \emph{OQ:} a typical OQ switch with a speedup of $N$.

\end{itemize}


\section{System Architecture}
\label{sec:system}

\subsection{Basic Crosspoint-Queued Switch}
\label{sec:basic}

\begin{figure}[ht]
\begin{minipage}[t]{3.2 in}
\centering \subfigure[The basic single-stage CQ switch.]{
\includegraphics[width=2.6in]{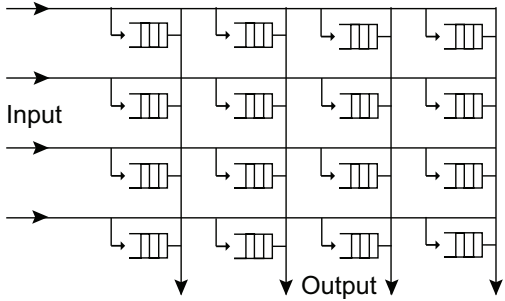}
\label{fig:cq}}
\end{minipage}
\begin{minipage}[t]{3.2 in}
\centering \subfigure[CCQ switch with a load balancer as the front stage.]{
\includegraphics[width=3.1in]{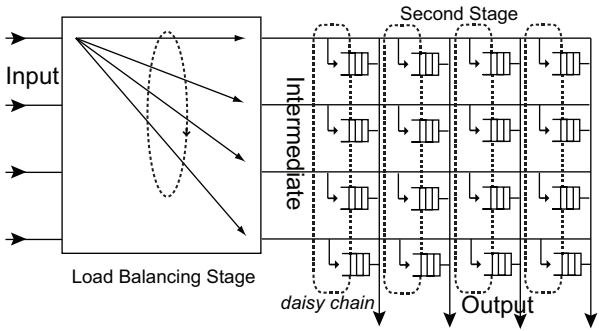}
\label{fig:lb}}
\end{minipage}
\caption{System architectures for crosspoint-queued switches.}
\label{fig:exampleIII}
\end{figure}

The single-chip CQ switch \cite{cqs} is a self-sufficient architecture which has all its buffers placed at the crosspoints of input and output lines, with no buffering at the input or output line-cards, as shown in Fig. \ref{fig:cq}.


Consider an $N\times N$ CQ switch with crosspoint buffers of size $B$ each, and let $0\le b_{ij}\le B$ denote the buffer occupancy at crosspoint $(i,j)$, $i,j=1,2,...,N$. The system is assumed to be time-slotted, in which packets are fragmented into fixed-length cells before entering the switch core. A header is also appended to each cell. Such headers may contain a cell ID, source/destination ports, etc. 

The basic \emph{CQ-LQF} scheduling scheme can be described as the following two phases in each time slot:

\begin{itemize}
\item
\textbf{\emph{Arrival Phase:}} For each input $i$, if there is a newly arriving cell destined to output $j$, it is directly sent to crosspoint $(i,j)$. If buffer $(i,j)$ is not full, i.e. $b_{ij}<B$, the new cell is accepted and buffered at the tail of line (TOL). Otherwise, this cell is dropped.

\item
\textbf{\emph{Departure Phase:}} For each output $j$, if not all crosspoints $(*,j)$ are empty, the output scheduler picks the one with the longest queue, and serves its HOL cell. If there are multiple longest queues of the same length, randomly pick one to break the tie.
\end{itemize}

The point of the LQF rule is that it always serves the fullest buffer that is the most likely to overflow. Since each output must determine the longest queue among all $N$ crosspoints in each time slot, its worst-case time complexity is at least $O(\log{N})$, assuming parallel comparator networks.

In this paper, we define that a cell belongs to flow $(i,j)$ if it travels from input $i$ to output $j$. For \emph{CQ-LQF}, cells of the same flow are always served in the same order as they arrive.

\subsection{Inefficiencies of Longest-Queue-First Scheduling}
\label{sec:inefficiency}

The basic CQ switch is simple and elegant. However, its buffers are small and segregated, which may result in a low buffer utilization and a high cell drop rate when the incoming traffic is bursty and non-uniform. The underlying reason is that such burstiness and non-uniformity may lead to unbalanced utilizations of these small buffers, even when the LQF rule is adopted. In this part, we show that LQF is not efficient enough for the CQ switch
. 

\subsubsection{Large Buffer Asymptotics}
\label{sec:largebuffer}
In studying the overflow probability or cell drop rate of queueing systems, much attention has been paid to their asymptotic behavior under the large buffer limit, and analysis of such large buffer asymptotics often relies on the theory of large deviations. Because counting the exact number of cell drops in various cases in a finite-buffer queueing system with general arrival processes is very complex and may not generate intuitive answers, we follow a common approach and turn to the approximate buffer overflow probability instead, i.e., the probability that the queue size $Q$ exceeds a certain value in an infinite-buffer queueing system.

The theory of large deviations is a powerful tool in the characterization of rare events like overflow in a queueing system. Let $\{X_t\}$ denote a stationary random arrival process, and $\{Y(t)\triangleq \sum_{\tau=1}^{t}{X_\tau}\}$ be the corresponding cumulative arrival process. For Bernoulli process, it is sufficient to use a single parameter $\lambda\triangleq E[X_t]$, the average arrival rate, to determine the process, i.e., $X_t(\lambda)=1$ with probability $\lambda$, and $X_t(\lambda)=0$ otherwise. Define $\Lambda^t(\theta,\lambda)=\log{E[e^{\theta Y_t(\lambda)}]}$ as the log moment generating function of the cumulative arrival process. 

According to \cite{bigqueues}, if the limit $\Lambda(\theta,\lambda)\triangleq \lim_{t\rightarrow \infty}{\Lambda^t(\theta,\lambda)/t}$ exists and is essentially smooth and finite in a neighborhood of $\theta=0$, then the stable queue size distribution of a single queue (SQ) with service rate $C$ under traffic $\{X_t\}$ satisfies
\begin{equation}
\label{eq:single01}
E^{\emph{SQ}}(C,\lambda)\triangleq \lim_{B\rightarrow \infty}{-\frac{1}{B}\log{P(Q>B)}}=\inf_{\gamma>0}{\gamma\Lambda^*(C+\frac{1}{\gamma},\lambda)},
\end{equation}
where $\Lambda^*(x,\lambda)\triangleq \sup_{\theta}{\theta x - \Lambda(\theta,\lambda)}$ is the convex conjugate or Fenchel-Legendre transform of $\Lambda(\theta,\lambda)$, and $E^{\emph{SQ}}(C,\lambda)$ is called the buffer overflow exponent of the SQ. The buffer overflow exponent is a function of $C$ and $\lambda$ for Bernoulli arrival processes, and represents the logarithmic decay rate of the overflow probability with respect to the buffer size. In other words, the higher the exponent, the faster the overflow probability drops given a certain amount of buffer increase.


Equation \ref{eq:single01} is the large buffer asymptotics for a single server queue fed by a single arrival process $\{X_t(\lambda)\}$. When $N$ i.i.d. processes are fed into a shared queue, the overall arrival process would be a superposition of these sources, i.e., $\Lambda^t_N(\theta,\lambda)=N\Lambda^t(\theta,\lambda)$. Correspondingly, $\Lambda_N(\theta,\lambda)=N\Lambda(\theta,\lambda)$, and hence $\Lambda^*_N(x,\lambda)\triangleq \sup_\theta{\theta x - \Lambda^t_N(\theta,\lambda)}=N(\sup_\theta{\theta \frac{x}{N}-\Lambda^t(\theta,\lambda)})=N\Lambda^*(\frac{x}{N},\lambda)$. This describes what happens at any output of an $N\times N$ OQ switch, and thus the buffer overflow exponent given uniform traffic arrival rate $\lambda$ per input-output pair and service rate $C$ per output would be 
\begin{equation}
\label{eq:oq}
\begin{aligned}
E_N^{\emph{OQ}}(C,\lambda) &\triangleq \lim_{B\rightarrow \infty}{-\frac{1}{B}\log{P(\sum_{i=1}^N{Q_i}>NB)}}\\
&=N\inf_{\gamma>0}{\gamma\Lambda_N^*(C+\frac{1}{\gamma},\lambda})\\
&=N^2\inf_{\gamma>0}{\gamma\Lambda^*(\frac{C+1/\gamma}{N},\lambda}),
\end{aligned}
\end{equation}
where $Q_i$ represents the queue size contributed by input $i$. In this paper, we set the service rate at each output to $C=1$, so the overflow exponent for \emph{OQ} is $E_N^{\emph{OQ}}(1,\lambda)$. Theorem $7$ in \cite{cqs} is an alternative expression of the large buffer asymptotics for the OQ switch.

On the other hand, for a CQ switch with LQF scheduling, theoretical analysis becomes much more complicated due to the separation of different queues. Here we leverage the analytical results done by Jagannathan et al. in a recent paper \cite{lqfoverflow}. Their main conclusion is that the buffer overflow exponent of $N$ separate queues with LQF scheduling can be expressed as that of an $n$-shared queueing system for some $n\le N$. More intuitively, in \emph{CQ-LQF}, when at least one crosspoint is full, only those crosspoints that are full at the same time get served by the output, while others are never served. Assume that there are $n\le N$ such crosspoints, then these crosspoints constitute a sub-system of mode $n$, which is equivalent to an OQ switch of the same size with the same arrival rate $\lambda$ and service rate $C$. The asymptotic buffer overflow performance of a CQ switch is determined by its lowest-performing mode, or the OQ sub-system with the highest overflow probability. The dominant overflow mode can be represented as a 4-tuple $(n_d,\lambda_d,C_d,B_d)$, where $n_d$ denotes the number of arrival processes fed into this OQ, $\lambda_d$ means the arrival rate to each arrival process, $C_d$ stands for the output service rate, and $B_d$ is the total buffer size of this OQ. A valid overflow mode should satisfy $n_d>C_d$.
\begin{equation}
\label{eq:cq}
\begin{aligned}
E_N^{\emph{CQ-LQF}}(C,\lambda) &\triangleq \lim_{B\rightarrow \infty}{-\frac{1}{B}\log{P(\max_{1\le i \le N}{Q_i}>B)}}\\
&=\min_{C<n\le N}{\lim_{B\rightarrow \infty}{-\frac{1}{B}\log{P(\sum_{i=1}^{n}{Q_i}>nB)}}}\\
&=\min_{C<n\le N}{E_n^{\emph{OQ}}(C,\lambda)}\\
&=\min_{C<n\le N}{n^2\inf_{\gamma>0}{\gamma\Lambda^*(\frac{C+1/\gamma}{n},\lambda})},
\end{aligned}
\end{equation}
where $n\ge C$ contains all possible overflow modes since otherwise the instantaneous arrival rate can never exceed the service rate, and $n^*\triangleq \arg\min_{n\in\mathcal{N}^*}{E_n^{\emph{OQ}}(1,\lambda)}$ determines the dominant mode $(n^*,\lambda,1,n^*B)$, which specifies the lowest-performing OQ sub-system with $n^*$ inputs, arrival rate $\lambda$ at each input, service rate $1$ at the output, and $n^*B$ buffers in all.

Another important corollary is that if the dominant $n$ is $n^*(\lambda)$, then it is most likely that $n^*(\lambda)$ out of $N$ queues will overflow together, while the other $N-n^*(\lambda)$ queues grow approximately to $n^*(\lambda)\gamma_{n^*}^*(\lambda)\lambda B$, where $\gamma_n^*(\lambda)\triangleq \arg \inf_{\gamma>0}{\gamma n\Lambda^*(\frac{1+1/\gamma}{n},\lambda)}$ is the optimal $t$ that achieves the infimum of $-E_n^{\emph{OQ}}(1,\lambda)$. In order to quantitatively compare the overflow performance in terms of buffer utilizations, we define a \emph{critical buffer utilization} $\eta$ as the expected overall utilization of all buffers upon overflow. For an OQ switch, $\eta_{\emph{OQ}}\equiv 100\%$. For a CQ switch, $\eta_{\emph{CQ}}=E(\frac{\sum_{i=1}^{N} Q_i}{NB}|\max_{1\le i \le N}{Q_i}>B)$.

According to Equation \ref{eq:cq}, it is obvious that the buffer overflow exponent for \emph{CQ-LQF} is no higher than that of \emph{OQ} of the same size, and may degenerate to \emph{OQ} of a smaller size $n\le N$ when the arrival rate is low. 

\begin{figure}[ht]
\centering
\begin{minipage}[t]{3.2 in}
\centering \subfigure[Buffer overflow exponent for \emph{OQ}.]{
\includegraphics[width=3.2 in]{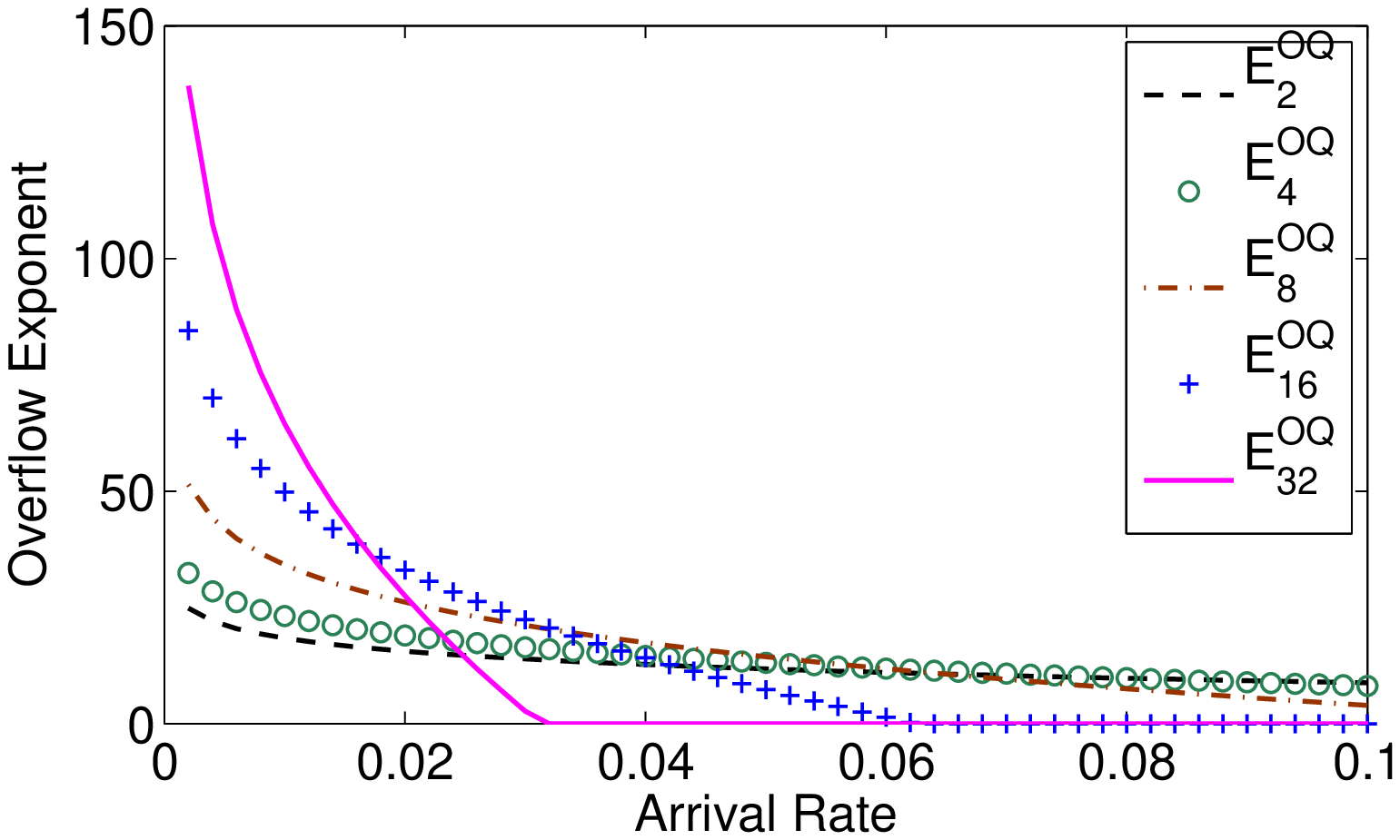}
\label{fig:EnOQ}}
\end{minipage}
\begin{minipage}[t]{3.2 in}
\centering \subfigure[Buffer overflow exponent \emph{CQ-LQF}.]{
\includegraphics[width=3.2 in]{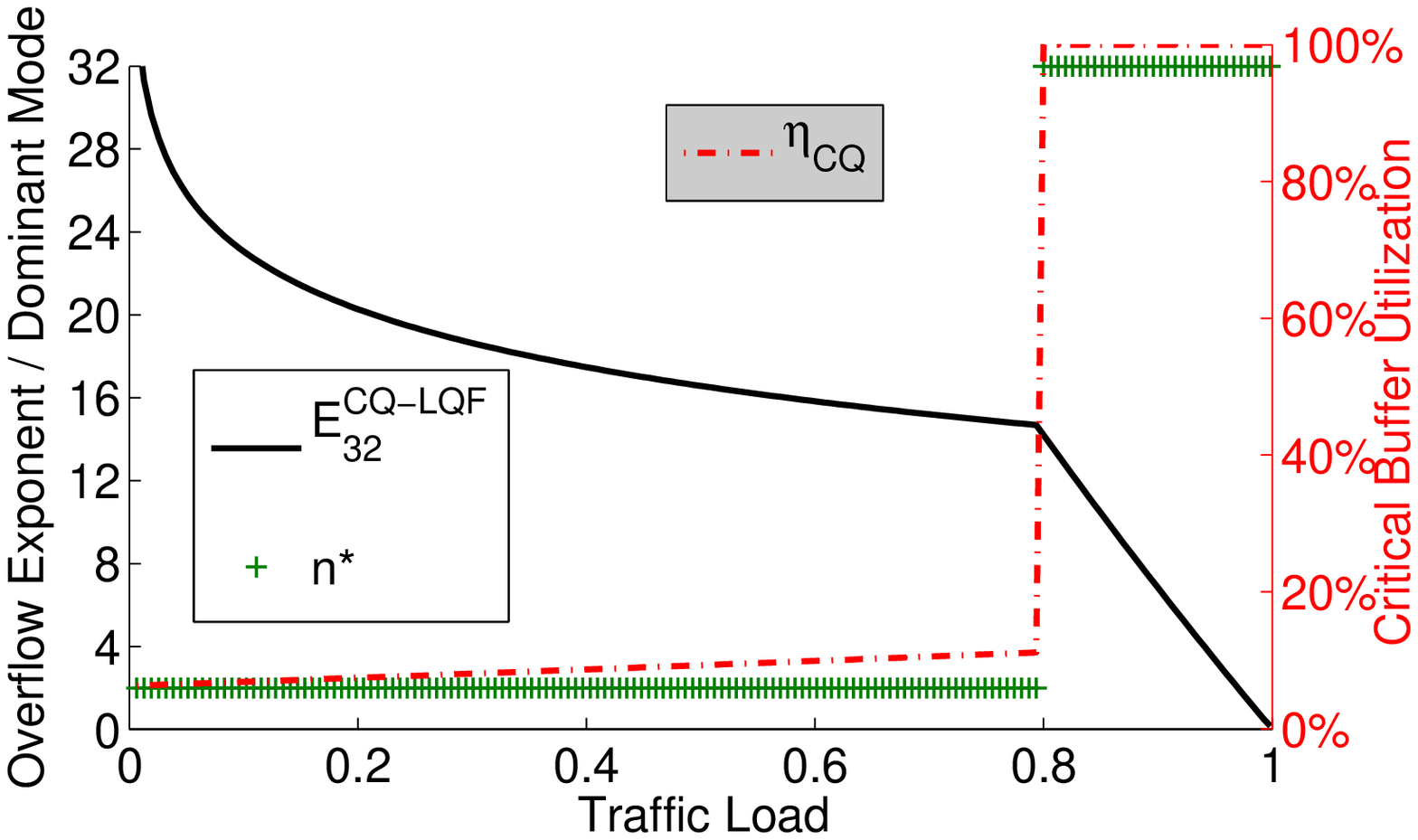}
\label{fig:E32LQF}}
\end{minipage}
\caption{Large buffer asymptotics for \emph{OQ} and \emph{CQ-LQF}.}
\label{fig:pooling}
\end{figure}

In Fig. \ref{fig:EnOQ}, we plot the buffer overflow exponents $E_n^{\emph{OQ}}(1,\lambda)$ for OQ switches of different sizes $n$, assuming uniform Bernoulli i.i.d. traffic across all inputs with $\Lambda(\theta,\lambda)=1-\lambda+\lambda e^\theta$. It can be seen that
\begin{itemize}
\item For fixed $n$, $E_n^{OQ}(1,\lambda)$ is monotonically decreasing with respect to $\lambda$, and drops to $0$ when $\lambda=1/n$;

\item For different $n$, $E_n^{OQ}(1,\lambda)$ with a larger $n$ starts at a higher value when $\lambda\rightarrow 0+$  but drops to $0$ faster.
\end{itemize}

In Fig. \ref{fig:E32LQF}, we take a CQ switch of size $N=32$ as an example, and show how $E_{32}^{\emph{CQ-LQF}}(1,\lambda)$ evolves with the normalized traffic load $\mu \triangleq N\lambda \in [0,1]$ at each output:
\begin{itemize}
\item For large $\mu \in [0.8,1]$ (or $0.025\le\lambda< 1/32$), $E_{32}^{\emph{CQ-LQF}}(1,\lambda)$ is determined by the characteristics of \emph{OQ} of the same size, and thus it is most likely that all queues would overflow at almost the same time, i.e., $n^*=N$;

\item For small $\mu\in (0.0.8)$ (or $\lambda<0.025$), $n^*=2$, and $E_{32}^{\emph{CQ-LQF}}(1,\lambda)$ abruptly degenerates to $E_{2}^{\emph{OQ}}(1,\lambda)$, so does the critical buffer utilization.
\end{itemize}

As we can see, compared with \emph{OQ}, \emph{CQ-LQF} cannot guarantee a high buffer utilization under low to medium traffic load, even if the large buffer limit is applied. Even though this does not mean \emph{CQ-LQF} always runs at such low buffer utilizations, we may still draw the conclusion that its buffer overflow performance is severely impaired by the separation of queues. The above results are derived under uniform traffic assumption, and the performance of the LQF policy can be even worse when traffic is non-uniform since some buffers may be consistently under-utilized and the longest queue might no longer be the most likely to overflow if its arrival rate is lower than others.


\subsubsection{Small Buffer Analysis}
In the previous part, we have revealed the inefficiency of LQF scheduling in the large buffer domain, which can be used only when uniform smooth traffic, e.g., Bernoulli i.i.d. traffic, is assumed given the typical buffer space in a CQ switch. Now we turn to its performance in the small buffer domain, which corresponds to more variable traffic sources that require larger buffer space than available.

We first investigate the impact of buffer size on the overflow exponent. According to Equation \ref{eq:single01}, when the arrival process is smooth, $\log{P(Q>B)}$ is linear with respect to $B$, which means adding a fixed amount of extra buffer always results in the same multiplicative decrease in the overflow probability, irrespective of the existing buffer size $B$ given that $B$ is sufficiently large. However, this is not true in the small buffer domain. Instead, Shwartz et al. \cite{small01} has reported that $\log{P(Q>B)}$ is proportional to $\sqrt{B}$ for small buffers fed by on/off sources with exponentially distributed sojourn times. The curve is essentially convex, and drops faster when $B$ is small. This result has later been confirmed and extended to more generally distributed arrival processes by Mandjes et al. \cite{smallbuffergeneral}. Moreover, for LRD traffic with $H\in (0.5, 1)$, $-\log{P(Q>B)}$ is always sub-linear with respect to $B$. Therefore, it is always more effective to increase the buffer size when $B$ is smaller. Thus the performance degradation for using \emph{CQ-LQF} rather than \emph{OQ} is significant.

The crosspoint buffer size is limited by the chip size and state-of-art ASIC technology. Compared with other legacy switches that may spread their buffer space on multiple chips, the buffer size of a CQ switch is still quite small, even as technological advances have eased this constraint. So the inefficient use of the segregated buffers needs to be addressed to improve the performance significantly. To make things worse, cells often arrive in bursts. With appropriate scalings, we know that a buffer of size $B$ facing bursts of fixed length $L$ has the same overflow exponent as a buffer of size $\frac{B}{L}$ facing Bernoulli i.i.d traffic. Thus buffer requirements increase when dealing with bursty traffic.

On the other hand, the LQF scheduling algorithm cannot perfectly balance the buffer utilizations, especially for small buffers fed by bursty and non-uniform traffic. In the worst case, a cell can be dropped as soon as two queues fill up, while all others are still empty. For a small switch with short buffers, if the arrival processes and the system state can be expressed as a Markov chain, then it is possible to derive the steady state probability distribtution and exact overflow probability or loss rate. In fact, Kanizo et al. \cite{cqs} has derived an exact expression for \emph{CQ-LQF} with buffer size $B=1$ under Bernoulli i.i.d. traffic, and their result could serve as an approximation for a general \emph{CQ-LQF} whose buffer size is comparable to the burst length.

\subsection{Combating Unbalanced Utilization}
\label{sec:modified}

Viewing the inefficiencies of the LQF policy, we apply efficient buffer sharing techniques to combat such unbalanced utilizations in the CQ switch.

\subsubsection{Load Balancing}
\label{sec:lb}

\begin{figure*}[ht]
\centering
\begin{minipage}[t]{2.3 in}
\centering \subfigure[Decomposing flows for $k$-distant crosspoints.]{
\includegraphics[width=2.3 in]{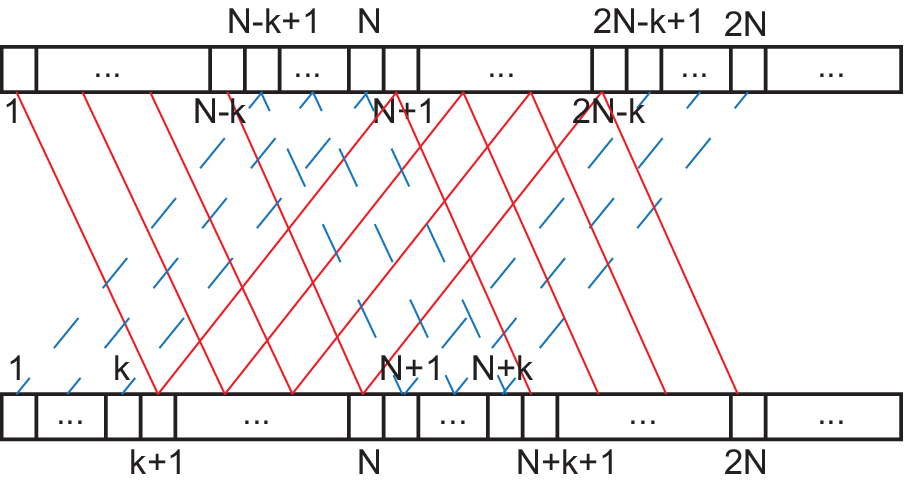}
\label{fig:kNk}}
\end{minipage}
\begin{minipage}[t]{2.3 in}
\centering \subfigure[Standard deviation versus distance.]{
\includegraphics[width=2.3 in]{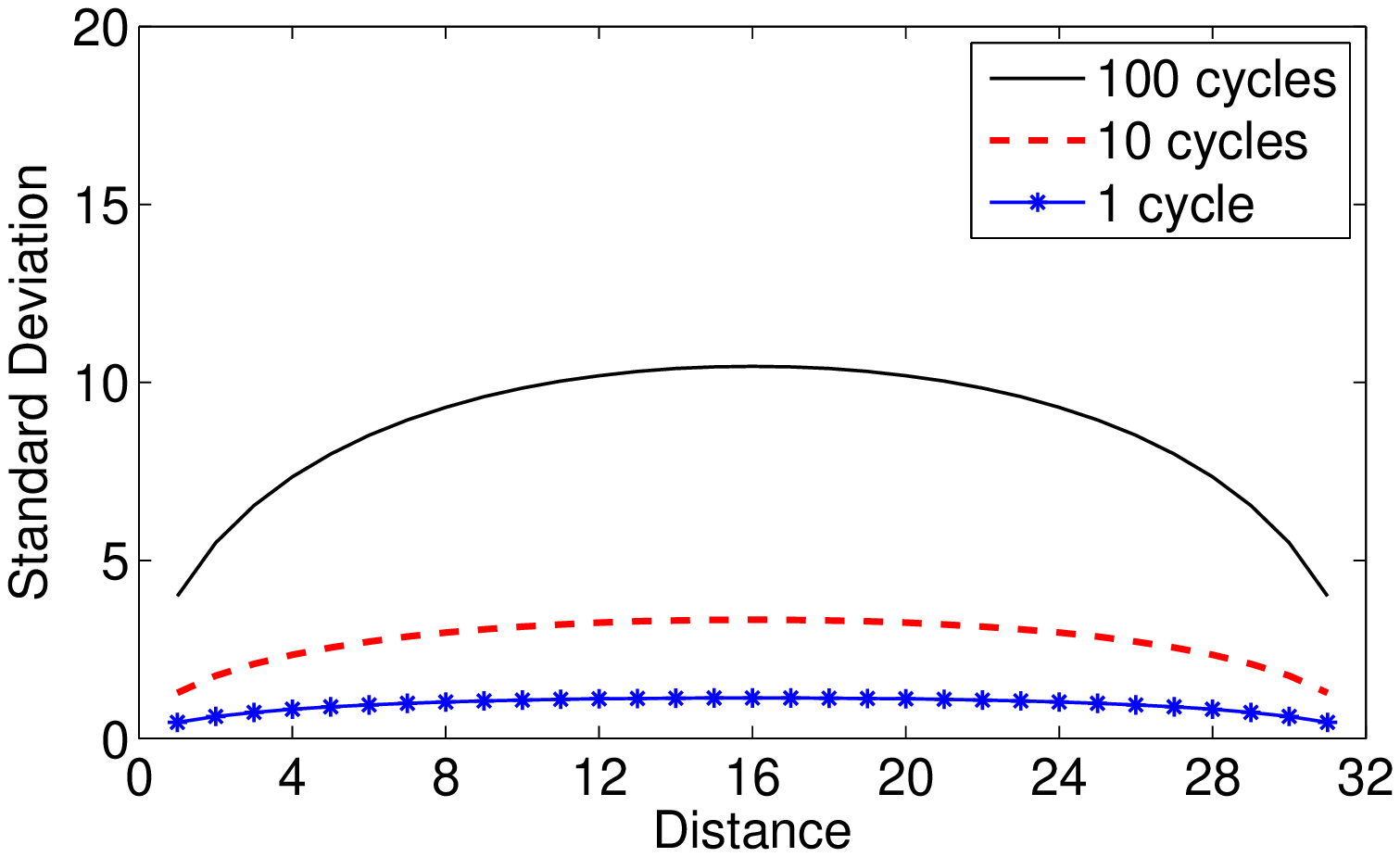}
\label{fig:distance}}
\end{minipage}
\begin{minipage}[t]{2.3 in}
\centering \subfigure[Standard deviation versus period.]{
\includegraphics[width=2.3 in]{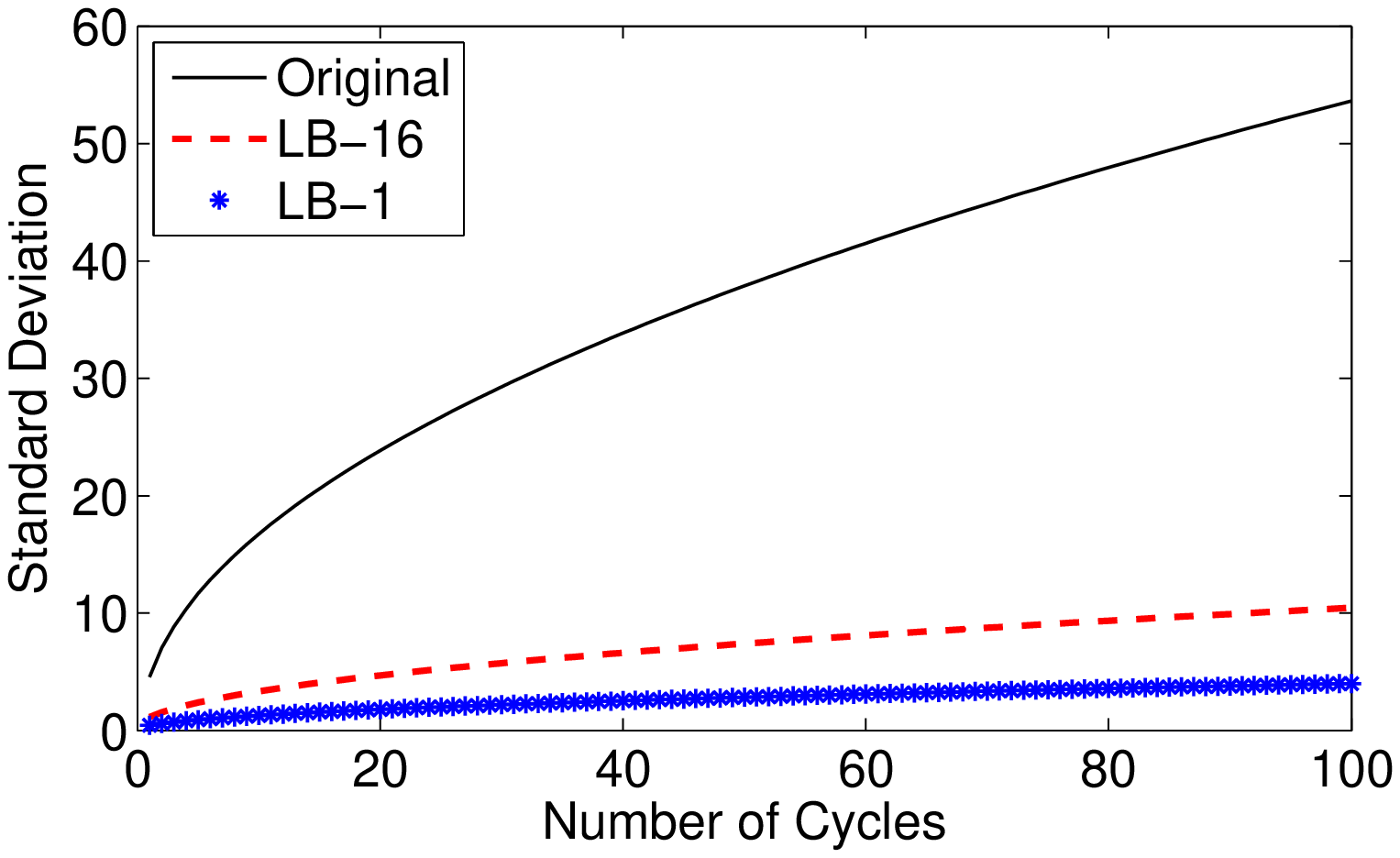}
\label{fig:standarddeviation}}
\end{minipage}
\caption{Variations in cumulative arrival processes under i.i.d. exponential ON-OFF traffic in a $32\times 32$ switch.}
\label{fig:difference}
\end{figure*}

First, we consider placing an extra load-balancer (first stage) in front of the CQ switching fabric (second stage), as shown in the left half of Fig. \ref{fig:lb} (the right half deals with the associated mis-sequencing problem, which will be presented in Section \ref{sec:RR}). The load-balancing stage walks through a fixed sequence of configurations: at time $t$, it connects each input $i$ to intermediate port $i+t$, which acts as both output $i+t$ of the first-stage and input $i+t$ of the second-stage. Effectively, $X^{LB}_{ij}(t)=X_{i-t,j}(t)$, where $X_{ij}(t)$ denotes the raw arrival process before load balancing and $X^{LB}_{ij}(t)$ is the arrival process after load balancing. Note that since the input and output port indices are always within $1$ through $N$, $i\pm t$ is an abbreviation of $mod(i\pm t-1,N)+1$. 


The load-balancer connects each input to each output in a round-robin fashion, and thus distributes the traffic equally to all crosspoints associated with the destination output. Let $\lambda_{i,j}$ denote the raw traffic arrival rate from input $i$ to output $j$, while $\lambda^{LB}_{i,j}$ represents the traffic arrival rate fed into crosspoint $(i,j)$ after passing the load-balancer, then
$
\lambda^{\emph{LB}}_{ij}=\frac{\sum_{i=1}^{N} \lambda_{ij}}{N}, \text{for } i,j=1,2,...,N.
$
In this way, the non-uniformity of the incoming traffic can be greatly reduced, since all crosspoint buffers $(i,j)$ associated with the same output $j$ essentially see the same arrival rate.

The load-balancer also effectively reduces the autocorrelation function of the traffic fed into any crosspoint $(i,j)$. Denote by $\rho (k)\triangleq Corr(X_{ij}(t),X_{ij}(t+k))$ the autocorrelation function of the raw incoming traffic at a lag of $k$ time-slots, and $\rho_0^{\emph{LB}}(k)\triangleq Corr(X^{\emph{LB}}_{ij}(t),X^{\emph{LB}}_{ij}(t+k))$ the autocorrelation function of the traffic after load balancing. 
Assume that the arrival processes are independent accross different inputs, then after passing the load-balancer, $\rho_0^{\emph{LB}}(k)=\rho(k)$ only at $k=0,\pm N, \pm 2N, \pm 3N...$, and $\rho_0^{\emph{LB}}(k)=0$ otherwise.
Therefore, the autocorrelation among consecutive arrivals is greatly suppressed by load balancing, thus reducing the burstiness of the incoming traffic.

At the same time, the arrivals to different queues now become correlated, which helps balance the cumulative arrival processes across different crosspoints,
$
\rho_k^{\emph{LB}}(k)\triangleq Corr(X^{\emph{LB}}_{ij}(t),X^{\emph{LB}}_{i+k,j}(t+k))=\rho(k).
$

Assuming uniform i.i.d. exponential ON-OFF traffic, we take a closer look at how load balancing affects the cumulative arrival processes to crosspoints associated with the same output. Now each flow follows the same Gilbert-Elliott 2-state Markov model, represented by state transition matrix
$\left[
\begin{array}{ccc}
 p_{00}& p_{01} \\
 p_{10}& p_{11}
\end{array}
\right]$, where 0 denotes the OFF state, 1 denotes the ON state, $p_{uv}$ represents the transition probability from state $u$ to state $v$ in one time-slot, and $\lambda=p_{01}/(p_{10}+p_{01})$.

The state transition matrix evolves over time as
\begin{equation}
\begin{aligned}
&\left[
\begin{array}{ccc}
 p_{00}(k)& p_{01}(k) \\
 p_{10}(k)& p_{11}(k)
\end{array}
\right] = \left[
\begin{array}{ccc}
 p_{00}& p_{01} \\
 p_{10}& p_{11}
\end{array}
\right]^k
\\&=\left[
\begin{array}{ccc}
\frac{p_{10}+p_{01}(1-p_{10}-p_{01})^k}{p_{10}+p_{01}}& \frac{p_{01}-p_{01}(1-p_{10}-p_{01})^k}{p_{10}+p_{01}} \\
\frac{p_{10}-p_{10}(1-p_{10}-p_{01})^k}{p_{10}+p_{01}}& \frac{p_{01}+p_{10}(1-p_{10}-p_{01})^k}{p_{10}+p_{01}}
\end{array}
\right].
\end{aligned}
\end{equation}

Without load balancing, the number of cells that arrive at each crosspoint during an arbitrary time period $t$ would be
\begin{equation}
Y(t)=
\begin{cases}
Y_1(t),&\text{with probability }\frac{p_{01}}{p_{10}+p_{01}}\\
Y_0(t),&\text{with probability }\frac{p_{10}}{p_{10}+p_{01}}
\end{cases},
\end{equation}
where $Y_u(t)$ represents the number of arrivals during period $(0,t]$ if the initial state is $u$ at time $0$:
\begin{equation}
Y_1(t)=
\begin{cases}
1+Y_1(t-1),&\text{with probability }p_{11}\\
1+Y_0(t-1),&\text{with probability }p_{10}
\end{cases},
\end{equation}
\begin{equation}
Y_0(t)=
\begin{cases}
Y_1(t-1),&\text{with probability }p_{01}\\
Y_0(t-1),&\text{with probability }p_{00}
\end{cases},
\end{equation}
with boundaries $Y_1(1)=1$ and $Y_0(1)=0$.

Let $\alpha\triangleq p_{11}-p_{01}$, and analyze the first moment of $Y(t)$,
\begin{equation}
\begin{aligned}
E[Y_1(t)]=1+\sum_{k=1}^{t-1}{p_{11}(k)}=\frac{p_{01}t}{1-\alpha}+\frac{p_{10}(1-\alpha^t)}{(1-\alpha)^2},
\end{aligned}
\end{equation}
\begin{equation}
E[Y_0(t)]=\sum_{k=1}^{t-1}{p_{01}(k)}=\frac{p_{01}t}{1-\alpha}-\frac{p_{01}(1-\alpha^t)}{(1-\alpha)^2},
\end{equation}
\begin{equation}
E[Y(t)]=\frac{p_{01}E[Y_1(t)]}{p_{10}+p_{01}}+\frac{p_{10}E[Y_1(t)]}{p_{10}+p_{01}}=\frac{p_{01}t}{p_{01}+p_{10}}.
\end{equation} 

Then we turn to the second moment,
\begin{equation}
\begin{aligned}
E[Y^2_1(t)]=&p_{11}(1+2E[Y_1(t-1)]+E[Y^2_1(t-1)])\\
&+p_{10}(1+2E[Y_0(t-1)]+E[Y^2_0(t-1)]),
\end{aligned}
\end{equation}
\begin{equation}
E[Y^2_0(t)]=p_{01}E[Y^2_1(t-1)]+p_{00}E[Y^2_0(t-1)],
\end{equation}

These are recursive formula. In order to derive explicit expressions, we need to view them in another way:
\begin{equation}
Y_1(t)=
\begin{cases}
1+Y_1(t-1),&\text{with probability }p_{11}\\
1+Y_1(t-2),&\text{with probability }p_{10}p_{01}\\
1+Y_1(t-3),&\text{with probability }p_{10}p_{00}p_{01}\\
...\\
1+Y_1(1),&\text{with probability }p_{10}p_{00}^{t-3}p_{01}\\
1,&\text{with probability }p_{11}
\end{cases}.
\end{equation}

$E[Y^2_1(1)]=1]$, $E[Y^2_1(2)]=1+3p_{11}$. For $t\ge 3$, we have
\begin{equation}
\begin{aligned}
E[Y^2_1(t)]=&p_{00}(E[Y^2_1(t-1)]-p_{11}E[(1+Y_1(t-2))^2])\\
&+p_{10}p_{01}E[(1+Y_1(t-2))^2]\\
&+p_{11}E[(1+Y_1(t-1))^2]\\
=&(1+\alpha)E[Y_1^2(t-1)]+2p_{11}E[Y_1(t-1)]\\
&-\alpha E[Y_1^2(t-2)]-2\alpha E[Y_1(t-2)]+p_{01}.
\end{aligned}
\end{equation}

Moving $E[Y^2_1(t-1)]$ to the left-hand side,
\begin{equation}
\begin{aligned}
&E[Y^2_1(t)]-E[Y^2_1(t-1)]\\
&=\alpha(E[Y_1^2(t-1)]-E[Y_1^2(t-2)])\\
&+2p_{11}E[Y_1(t-1)]-2\alpha E[Y_1(t-2)]+p_{01}.
\end{aligned}
\end{equation}

Replacing $E[Y_1^2(t-1)]-E[Y_1^2(t-2)]$ recursively, 
we have
\begin{equation}
\begin{aligned}
&E[Y^2_1(t)]-E[Y^2_1(t-1)]=c_1\frac{1-\alpha^{t-2}}{1-\alpha}+c_2(t-2)\alpha^{t-2}\\
&+c_3\left(\frac{t}{1-\alpha}-\frac{\alpha}{(1-\alpha)^2}+\frac{3\alpha-2}{(1-\alpha)^2}\alpha^{t-2}\right)+3p_{11}\alpha^{t-1},
\end{aligned}
\end{equation}
where $c_1\triangleq \frac{p_{01}p_{11}-3p_{01}^2+p_{01}}{1-\alpha}+\frac{2p_{01}p_{10}}{(1-\alpha)^2}$, $c_2\triangleq \frac{2p_{10}^2\alpha}{(1-\alpha)^2}$, and $c_3\triangleq \frac{2p_{01}^2}{1-\alpha}$. Then summing these items up for $t\ge 2$,
\begin{equation}
\begin{aligned}
E[Y^2_1(t)]=&\sum_{\tau=3}^{t}{(E[Y^2_1(\tau)]-E[Y^2_1(\tau-1)])}+E[Y^2_1(2)]\\
=&c_4+c_5t+c_6t^2+c_7\alpha^{t-1}+c_8t\alpha^{t-1},
\end{aligned}
\end{equation}
where $c_4\triangleq 1+3p_{11}+\frac{c_3\alpha(3\alpha-2)}{(1-\alpha)^3}+\frac{3p_{11}\alpha-2c_1-3c_3-2c_2\alpha}{1-\alpha}+\frac{2c_3\alpha+c_2\alpha(3-2\alpha)-c_1\alpha}{(1-\alpha)^2}$, $c_5\triangleq \frac{2c_1+c_3}{2(1-\alpha)}-\frac{c_3\alpha}{(1-\alpha)^2}$, $c_6\triangleq \frac{c_3}{2(1-\alpha)}$, $c_7\triangleq \frac{2c_2-3p_{11}}{1-\alpha}+\frac{c_1-c_2}{(1-\alpha)^2}-\frac{c_3(3\alpha-2)}{(1-\alpha)^3}$, and $c_8\triangleq -\frac{c_2}{1-\alpha}$.

Similarly, $E[Y_0^2(1)]=0$, and for $t\ge 2$,
\begin{equation}
Y_0(t)=
\begin{cases}
Y_1(t-1),&\text{with probability }p_{01}\\
Y_1(t-2),&\text{with probability }p_{00}p_{01}\\
Y_1(t-3),&\text{with probability }p_{00}^2p_{01}\\
...\\
Y_1(1),&\text{with probability }p_{00}^{t-2}p_{01}\\
0,&\text{with probability }p_{00}^{t-1}
\end{cases}.
\end{equation}
\begin{equation}
\begin{aligned}
E[Y^2_0(t)]=&\sum_{\tau=1}^{t-1}{p_{00}^{t-1-\tau}p_{01}E[Y^2_1(\tau)]}\\
=&c_4\sum_{\tau=1}^{t-2}{p_{01}p_{00}^{\tau-1}}+c_5\sum_{\tau=1}^{t-2}{p_{01}p_{00}^{\tau-1}(t-\tau)}\\
&+c_6\sum_{\tau=1}^{t-2}{p_{01}p_{00}^{\tau-1}(t-\tau)^2}+c_7\sum_{\tau=1}^{t-2}{p_{01}p_{00}^{\tau-1}\alpha^{t-\tau}}\\
&+c_8\sum_{\tau=1}^{t-2}{p_{01}p_{00}^{\tau-1}(t-\tau)\alpha^{t-1-\tau}}+p_{01}p_{00}^{t-2}.
\end{aligned}
\end{equation}

Then we calculate the variance of $Y(t)$,
\begin{equation}
\begin{aligned}
\sigma_Y^2(t)=&E[Y^2(t)]-E^2[Y(t)]\\
=&\frac{p_{01}E[Y^2_1(t)]}{p_{10}+p_{01}}+\frac{p_{10}E[Y^2_0(t)]}{p_{10}+p_{01}}-E^2[Y(t)]\\
=&\frac{p_{01}}{1-\alpha}(c_4+c_5t+c_6t^2+c_7\alpha^{t-1}+c_8t\alpha^{t-1})\\
&+\frac{p_{10}}{1-\alpha}\left(c_4\sum_{\tau=1}^{t-2}{p_{01}p_{00}^{\tau-1}}+c_5\sum_{\tau=1}^{t-2}{p_{01}p_{00}^{\tau-1}(t-\tau)}\right. \\
&+c_6\sum_{\tau=1}^{t-2}{p_{01}p_{00}^{\tau-1}(t-\tau)^2}+c_7\sum_{\tau=1}^{t-2}{p_{01}p_{00}^{\tau-1}\alpha^{t-\tau}}\\
&\left.+c_8\sum_{\tau=1}^{t-2}{p_{01}p_{00}^{\tau-1}(t-\tau)\alpha^{t-1-\tau}}+p_{01}p_{00}^{t-2}\right)\\
&-\frac{p_{01}^2t^2}{(1-\alpha)^2}.
\end{aligned}
\end{equation} 
There is a similar analysis of the variance-time curve in \cite{variancetime}, but the OFF period in that paper is not slotted.

Since the arrival processes fed into each crosspoint are i.i.d., the difference in cumulative arrivals between any two crosspoints $\Delta_{\emph{orig}}(t) \triangleq Y(t)-Y'(t)$ should satisfy $E[\Delta_{\emph{orig}}(t)]=0$ and $\sigma_{\emph{orig}}^2(t)=2\sigma_Y^2(t)$.

On the other hand, with load balancing, the arrival processes to each crosspoint associated with the same output are correlated. The conditional probability of crosspoint $i+k$ being in state $v$ at time $k$ given crosspoint $i$ being in state $u$ at time $0$ is $p_{uv}(k)$ because both cells belong to the same flow from input $i$ before being load balanced. Meanwhile, the conditional probability of crosspoint $i$ being in state $w$ at time $N$ given crosspoint $i+k$ being in state $v$ at time $k$ is $p_{vw}(N-k)$. Fixing this specific $(k,N-k)$-periodic flow $i$, and analyzing its contribution to the difference of cumulative arrivals between two $k$-distant crosspoints during $t$ cycles of $N$ time-slots each, we have
\begin{equation}
\delta_k(t)=
\begin{cases}
\delta_{1k}(t),&\text{with probability }\frac{p_{01}}{p_{10}+p_{01}}\\
\delta_{0k}(t),&\text{with probability }\frac{p_{10}}{p_{10}+p_{01}}
\end{cases},
\end{equation}
where $\delta_{uk}(t)$ denotes the difference in cumulative arrivals contributed by flow $i$ alone during time $(0,Nt]$ if the initial state is $u$ at crosspoint $i$ at time $0$:
\begin{equation}
\delta_{1k}(t)=
\begin{cases}
\delta_{1k}(t-1),&\text{with }p_{11}(k)p_{11}(N-k)\\
\delta_{0k}(t-1),&\text{with }p_{11}(k)p_{10}(N-k)\\
-1+\delta_{1k}(t-1),&\text{with }p_{10}(k)p_{01}(N-k)\\
-1+\delta_{0k}(t-1),&\text{with }p_{10}(k)p_{00}(N-k)
\end{cases},
\end{equation}
\begin{equation}
\delta_{0k}(t)=
\begin{cases}
\delta_{1k}(t-1),&\text{with }p_{00}(k)p_{01}(N-k)\\
\delta_{0k}(t-1),&\text{with }p_{00}(k)p_{00}(N-k)\\
1+\delta_{1k}(t-1),&\text{with }p_{01}(k)p_{11}(N-k)\\
1+\delta_{0k}(t-1),&\text{with }p_{01}(k)p_{10}(N-k)
\end{cases},
\end{equation}
with boundaries $\delta_{uk}(0)=0$ for any $u$ and $k$.

Calculating the first moment of $\delta_k(Nt)$,
\begin{equation}
E[\delta_{1k}(t)]=-\frac{p_{10}(1-\alpha^k)(1-\alpha^{Nt})}{(1-\alpha)(1-\alpha^N)},
\end{equation}
\begin{equation}
E[\delta_{0k}(t)]=\frac{p_{01}(1-\alpha^k)(1-\alpha^{Nt})}{(1-\alpha)(1-\alpha^N)},
\end{equation}
\begin{equation}
E[\delta_{k}(t)]=\frac{p_{01}E[\delta_{1k}(t)]}{p_{10}+p_{01}}+\frac{p_{10}E[\delta_{0k}(t)]}{p_{10}+p_{01}}=0.
\end{equation}

In terms of the second moment,
\begin{equation}
\begin{aligned}
E[\delta^2_{1k}(t)]=&p_{10}(N)E[\delta^2_{0k}(t-1)]+p_{11}(N)E[\delta^2_{1k}(t-1)]\\
&+p_{10}(k)-2p_{10}(k)p_{00}(N-k)E[\delta_{0k}(t-1)]\\
&-2p_{10}(k)p_{01}(N-k)E[\delta_{1k}(t-1)]\\
=&\frac{2p_{01}p_{10}(1-\alpha^k)(1-\alpha^{N-k})t}{(1-\alpha)^2(1-\alpha^N)}\\
&-\frac{2p_{01}p_{10}(1-\alpha^k)(1-\alpha^{N-k})(1-\alpha^{Nt})}{(1-\alpha)^2(1-\alpha^N)^2}\\
&+\frac{p_{10}(1-\alpha^{Nt})(1-\alpha^k)}{(1-\alpha)(1-\alpha^N)},
\end{aligned}
\end{equation}
\begin{equation}
\begin{aligned}
E[\delta^2_{0k}(t)]=&p_{00}(N)E[\delta^2_{0k}(t-1)]+p_{01}(N)E[\delta^2_{1k}(t-1)]\\
&+p_{01}(k)+2p_{01}(k)p_{10}(N-k)E[\delta_{0k}(t-1)]\\
&+2p_{01}(k)p_{11}(N-k)E[\delta_{1k}(t-1)]\\
=&\frac{2p_{01}p_{10}(1-\alpha^k)(1-\alpha^{N-k})t}{(1-\alpha)^2(1-\alpha^N)}\\
&-\frac{2p_{01}p_{10}(1-\alpha^k)(1-\alpha^{N-k})(1-\alpha^{Nt})}{(1-\alpha)^2(1-\alpha^N)^2}\\
&+\frac{p_{01}(1-\alpha^{Nt})(1-\alpha^k)}{(1-\alpha)(1-\alpha^N)},
\end{aligned}
\end{equation}
\begin{equation}
\begin{aligned}
E[\delta^2_k(t)]=&\frac{p_{01}E[\delta^2_{1k}(t)]}{p_{10}+p_{01}}+\frac{p_{10}E[\delta^2_{0k}(t)]}{p_{10}+p_{01}}\\
=&\frac{2p_{01}p_{10}(1-\alpha^k)(1-\alpha^{N-k})t}{(1-\alpha)^2(1-\alpha^N)}\\
&+\frac{2p_{01}p_{10}\alpha^{N-k}(1-\alpha^k)^2(1-\alpha^{Nt})}{(1-\alpha)^2(1-\alpha^N)^2},
\end{aligned} 
\end{equation}
\begin{equation}
\begin{aligned}
\sigma_{k}^2(t)=&E[\delta_k^2(t)]-E^2[\delta_k(t)]\\
=&\frac{2p_{01}p_{10}(1-\alpha^k)(1-\alpha^{N-k})t}{(1-\alpha)^2(1-\alpha^N)}\\
&+\frac{2p_{01}p_{10}\alpha^{N-k}(1-\alpha^k)^2(1-\alpha^{Nt})}{(1-\alpha)^2(1-\alpha^N)^2}.
\end{aligned}
\end{equation} 

Summing up all $N$ i.i.d. flows during period $(0,Nt]$ as in Fig. \ref{fig:kNk}, the overall difference in cumulative arrivals equals
\begin{equation}
\Delta_{\emph{LB-}k}(Nt)= \sum_{i=1}^{N-k}{\delta_k^{(i)}(t)}-\sum_{i=1}^{k}{\delta_{N-k}^{(i)}(t)},
\end{equation}
thus its first and second moments are
\begin{equation}
E[\Delta_{\emph{LB-}k}(Nt)]=(N-k)E[\delta_k(t)]-kE[\delta_{N-k}(t)]=0,
\end{equation}
\begin{equation}
\begin{aligned}
\sigma^2_{\emph{LB-}k}(Nt)=&(N-k)\sigma^2_k(t)+k\sigma^2_{N-k}(t)\\
=&\frac{2Ntp_{01}p_{10}(1-\alpha^k)(1-\alpha^{N-k})}{(1-\alpha)^2(1-\alpha^N)}\\
&+\frac{2p_{01}p_{10}(1-\alpha^{Nt})}{(1-\alpha)^2(1-\alpha^N)^2}\\
&\times((N-k)\alpha^{N-k}(1-\alpha^k)^2+k\alpha^{k}(1-\alpha^{N-k})^2).
\end{aligned}
\end{equation}

From the expression above, it can be found that $\sigma^2_{\emph{LB-}k}(Nt)$ is symmetric about $k=\frac{N}{2}$. Meanwhile, $\frac{\partial \sigma^2_{\emph{LB-}k}(Nt)}{\partial k}=0$ at $k=\frac{N}{2}$, and $\frac{\partial \sigma^2_{\emph{LB-}k}(Nt)}{\partial k}>0$ for large $t$ and $\forall k\in(0,N)$. Therefore, the curve is concave and unimodal at $k=\frac{N}{2}$.

For illustration purposes, we consider a $32\times 32$ CQ switch fed by exponential ON-OFF traffic with state transition matrix $\left[
\begin{array}{ccc}
 p_{00}& p_{01} \\
 p_{10}& p_{11}
\end{array}
\right]=\left[
\begin{array}{ccc}
 389/390& 1/390 \\
 0.1& 0.9
\end{array}
\right]$. In Fig. \ref{fig:distance}, we plot how $\sigma_{\emph{LB-}k}$ changes with the distance between two crosspoints over various time periods. It is verified that the standard difference is symmetric about and reaches its maximum at $k=16$ regardless of the time period (at least $N$ time-slots). This means crosspoints that are close together (in either directions) tend to have similar arrivals. In addition, we also compare $\sigma_{\emph{orig}}$ with $\sigma_{\emph{LB-}k}$ in Fig. \ref{fig:standarddeviation}. As time passes, the variation of cumulative arrival processes grows sub-linearly. Meanwhile, load balancing dramatically suppresses such variations by transforming independent, bursty arrivals into correlated, less-bursty ones.

\subsubsection{Deflection Routing}
\label{sec:deflection}

We also consider deflection routing to actively balance the buffer utilizations of different crosspoints, and develop an augmented architecture, the CCQ switch, which is suitable for deflection routing (and packet ordering for load balancing) when combined with the scheduling schemes to be proposed in Section \ref{sec:CCQ}.

In the CCQ switch, crosspoints associated with a common output port are single-connected into a daisy chain (in the order of their associated input port indices), as shown in Fig. \ref{fig:lb}. Specifically, crosspoint $(i,j)$ is connected with its predecessor $(i-1,j)$ and successor $(i+1,j)$.

With this modification, cell deflection (and message passing) can be easily supported between adjacent crosspoints along the daisy chains. In terms of the hardware requirement, by adding an extra layer of connections, we introduce an extra memory-read speedup and an extra memory-write speedup for each crosspoint buffer. The extra speedup and inter-crosspoint connections are purely internal to the switch core, implemented on a single chip, and thus do not impose extra burdens on the links between the input/output line-cards and the switching core (card-edge and chip-pin limitations \cite{asic03}).

The main idea of deflection routing is to reroute cells from highly-occupied crosspoints to their less-occupied neighbors, just like water flows from a higher elevation to a lower elevation. Compared with the LQF policy, deflection routing is usually more effective in reducing unbalanced loads. Multiple deflections from highly utilized crosspoints to their under-utilized neighbors can occur in one time slot, while LQF can only reduce the length of one queue in each time slot. Also, unlike load balancing, deflection routing is a reactive strategy which redistributes incoming cells after they have already flooded into the crosspoints. Ideally, given enough time with no new arrivals or departures, the buffer utilization of all crosspoints in the same daisy chain can be perfectly equalized. In this sense, deflection routing can alleviate the problem associated with LQF in the large buffer domain. Moreover, deflection routing and load balancing are complementary to each other, and can be combined to work together so that any unbalances can be further suppressed.

As will be shown in Section \ref{sec:CCQ}, load balancing and deflection routing can both be supported on a two-stage CCQ switch. However, their synergy might not be fully exploited if the interactions between these two techniques are overlooked.

In fact, we have been insisting that load balancing should follow a strict order of $1\rightarrow 2\rightarrow3\rightarrow...\rightarrow N$, while deflection routing is also restricted to neighboring crosspoints $N\rightarrow N-1 \rightarrow...\rightarrow 1$. Such a similarity has a side effect when the two techniques are combined together, that is, the correlated arrivals incurred by load balancing may impair the effectiveness of deflection routing. To be more specific, we focus on two arbitrary neighboring crosspoints $(i-1,j)$ and $(i,j)$. The load-balanced arrival processes that are fed into these two crosspoints satisfy $\rho_1^{\emph{LB}}(1)= Corr(X^{\emph{LB}}_{i-1,j}(t-1),X^{\emph{LB}}_{i,j}(t))=\rho(1)$. If the original arrival process is highly bursty and self-correlated, then there are strong correlations between the load-balanced arrival processes as well. Similar correlations also exist between the departure processes, and hence the buffer occupancies as well. This correlation means that deflection routing will be relatively ineffective if applied to two neighboring buffers, since deflection routing exploits differences in buffer occupancy.

This problem can be solved by disrupting the consistency between load balancing and deflection routing orders. As shown in Fig. \ref{fig:distance}, neighboring crosspoints have minimal difference in cumulative arrivals, while crosspoints that are $\frac{N}{2}$ distance away may be least correlated. Therefore, if we let deflection routing take place between crosspoints that are far away during load balancing, more fluctuations in buffer occupancies can be expected throughout the daisy chain, and thus deflection routing will have more opportunities to balance the buffer utilizations locally and reach a global equilibrium faster. This will be discussed in further detail in Section \ref{sec:CCQ} after the scheduling schemes for CCQ switches are proposed.

\subsubsection{Buffer Pooling}
\label{sec:wxr}

\begin{figure*}[ht]
\centering
\begin{minipage}[t]{2.3 in}
\centering \subfigure[$4\times 1$ pooling pattern.]{
\includegraphics[width=2.2 in]{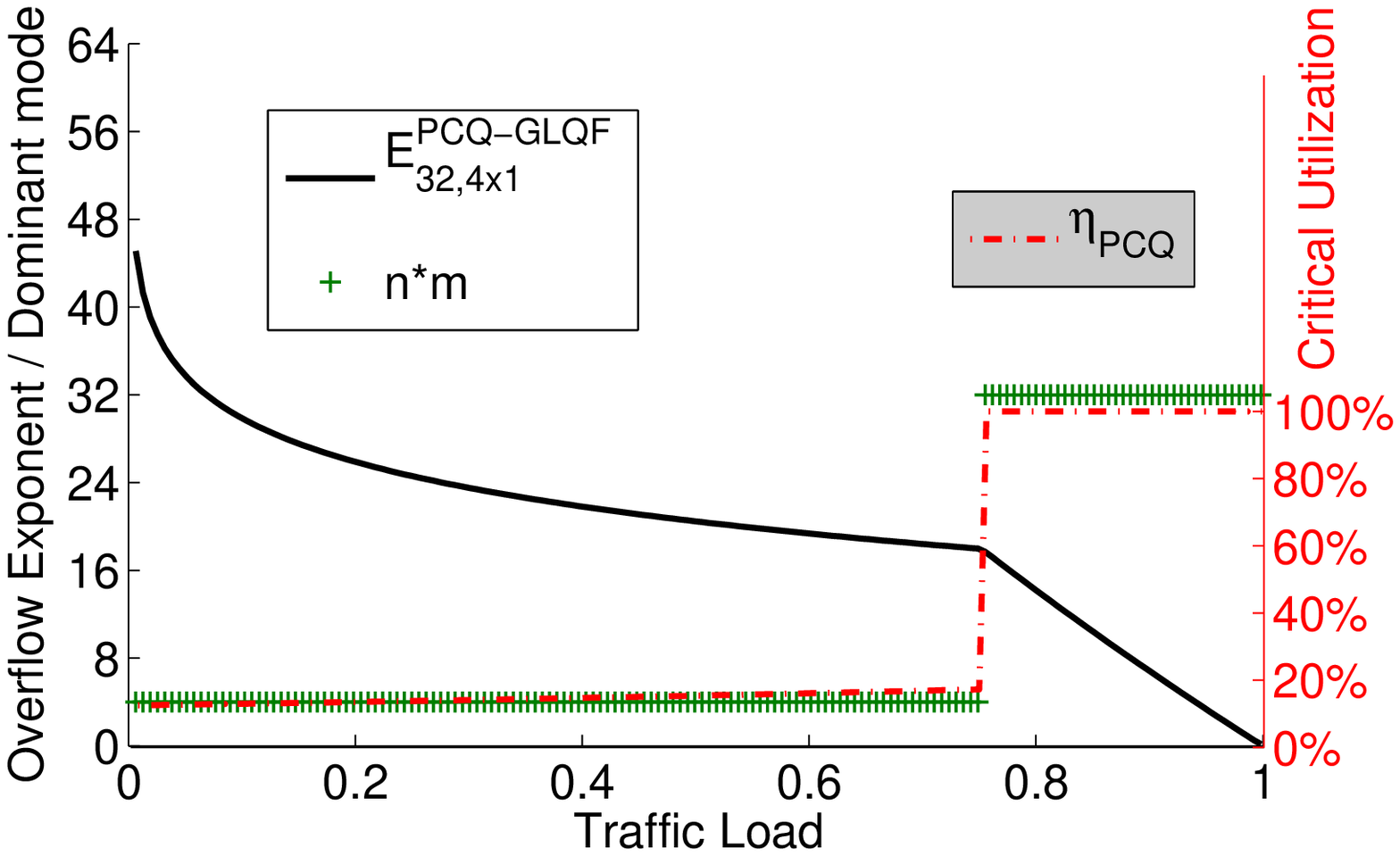}
\label{fig:mx1}}
\end{minipage}
\begin{minipage}[t]{2.3 in}
\centering \subfigure[$1\times 4$ pooling pattern.]{
\includegraphics[width=2.2 in]{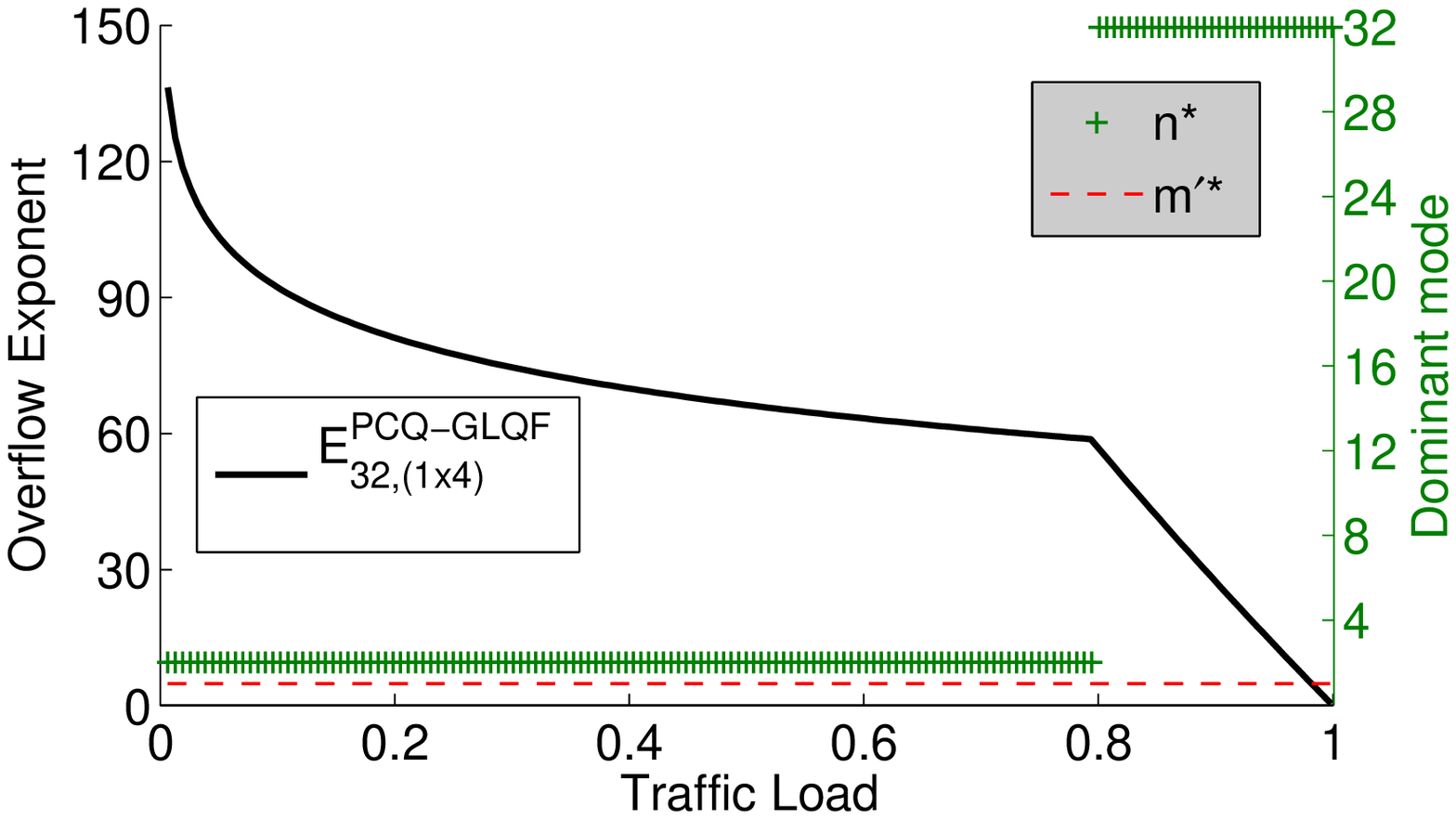}
\label{fig:1xm}}
\end{minipage}
\begin{minipage}[t]{2.3 in}
\centering \subfigure[$2\times 2$ pooling pattern.]{
\includegraphics[width=2.2 in]{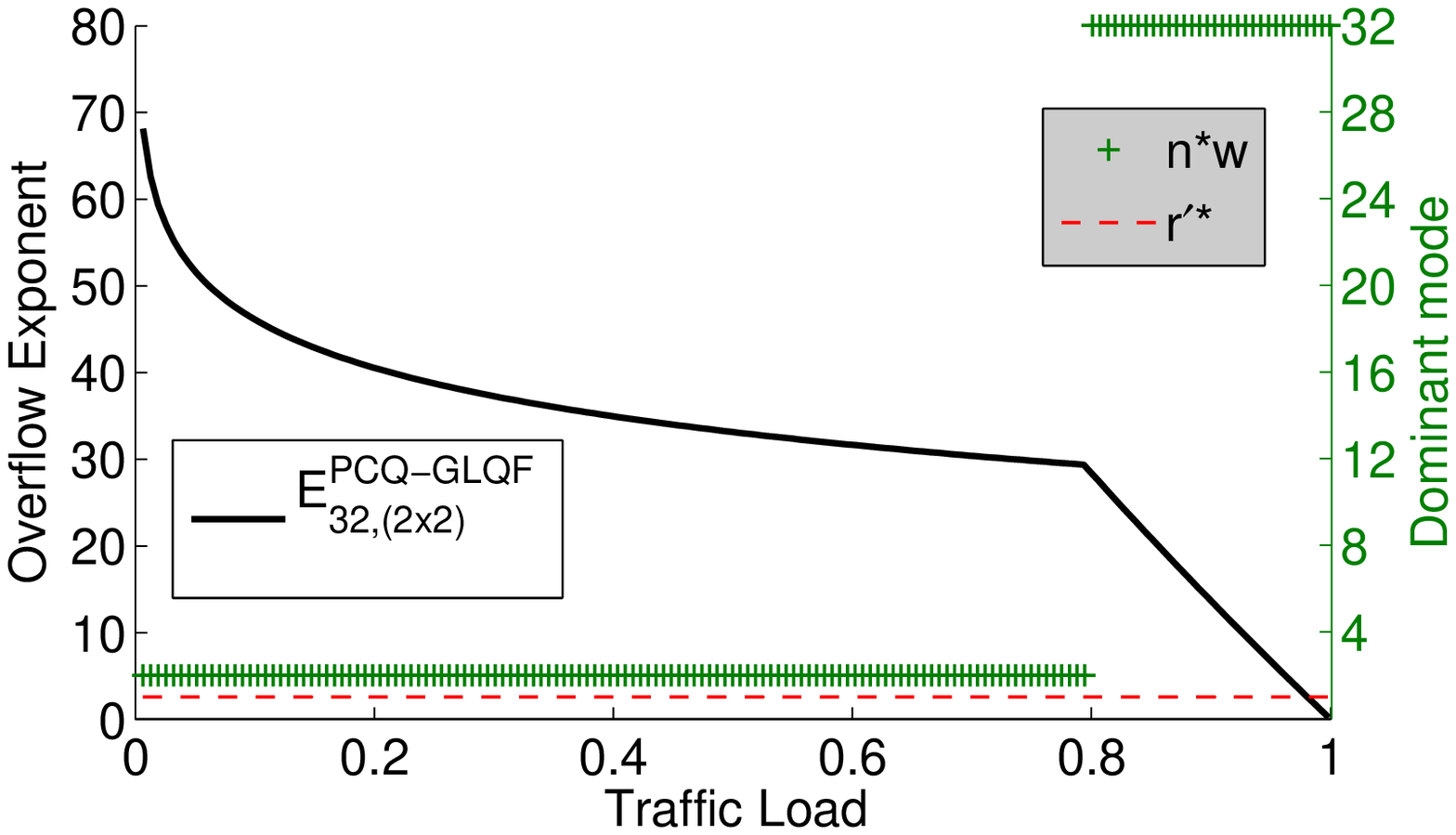}
\label{fig:wxr}}
\end{minipage}
\caption{Large buffer asymptotics for \emph{PCQ-GLQF}.}
\label{fig:pooling}
\end{figure*}

The CQ switch benefits from the flexibility of routing and ease of scheduling facilitated by the mesh-connected crosspoint buffers, but it also suffers from the low utilization caused by fragmentation of the limited buffer space. It is quite natural to think of pooling buffers together, and there are some tradeoffs between flexibility, complexity, utilization and speedup. Specifically, we can use a larger buffer to serve multiple crosspoints instead of a smaller buffer for each crosspoint, so that buffer space can be dynamically allocated among busy and idle crosspoints, at the cost of some extra hardware memory speedup and/or scheduling complexity.

In principle, any $m$ crosspoints can be aggregated together to form a buffer pool, and such $m$ can vary across different pools. However, we restrict ourselves to a class of $w\times r$ rectangular pooling patterns for ease of analysis in this paper. 
A memory-write speedup of $1\le s_w\le w$ and memory-read speedup of $1\le s_r\le r$ are assumed to resolve input and output contentions.

Under $w\times r$ pooling and uniform Bernoulli i.i.d. traffic of rate $\lambda\le \frac{1}{N}$, the probability that $k\ge 1$ crosspoints in the same pool receive cells at the same time is 
$
\label{eq:mwr}
P_{w\times r}^{k}=\binom{w}{k}(r\lambda)^k(1-r\lambda)^{w-k}\le (m\lambda)^k \le \left(\frac{m}{N}\right)^k,
$
so it is very unlikely that many crosspoints will receive cells at the same time if $m\ll N$, and thus the aggregation (and dynamic allocation) of these crosspoints ``virtually'' increases the amount of buffers that can be used by each crosspoint without extending the actual buffer space. When $m$ crosspoints are pooled together, the effective buffer size seen by each crosspoint almost grows linearly if $m\ll N$, as we will illustrate next. Considering the convex loss curve for small buffer and/or LRD traffic, such a multiplexing gain is especially crucial in the small buffer domain of CQ switches.

Next, we show how buffer pooling may affect the overflow probability. For simplicity, we assume an ideal generalized longest-queue-first (GLQF) policy. Under this policy, every output takes turns to reserve a cell from the most occupied buffer pool and update the remaining occupancy of that pool. Then the buffer pools are served according to those reservations simultaneously, regardless of the speedup requirements.

We first consider $m\times 1$ pooling patterns. When $n'$ such pooled buffers constitute an OQ switch, $E_{n',(m\times 1)}^{\emph{OQ}}(C,\lambda) \triangleq \lim_{B\rightarrow \infty}{-\frac{1}{B}\log{P(\sum_{i=1}^{n'm}{Q_i}>n'mB)}}
=E_{n'm}^{\emph{OQ}}(C,\lambda)$.
Then consider a PCQ switch of size $N$ with LQF scheduling. Following the same approach as in Equation \ref{eq:cq}, its buffer overflow exponent could be expressed as
\begin{equation}
\label{eq:mx1}
\begin{aligned}
&E_{N,(m\times 1)}^{\emph{PCQ-GLQF}} (C,\lambda)\\
&\triangleq  \lim_{B\rightarrow \infty}{-\frac{1}{B}\log{P(\max_{1\le I \le \frac{N}{m}}{\sum_{i=(I-1)m+1}^{Im}{Q_i}}>mB)}}\\
&= \min_{\frac{C}{m}<n\le\frac{N}{m}}{\lim_{B\rightarrow \infty}{-\frac{1}{B}\log{P(\sum_{I=1}^{n}{\sum_{i=(I-1)m+1}^{Im}{Q_i}}>nmB)}}}\\
&= \min_{\frac{C}{m}<n\le\frac{N}{m}}{E_{n,(m\times 1)}^{\emph{OQ}}(C,\lambda)}
=\min_{\frac{C}{m}<n\le\frac{N}{m}}{E_{nm}^{\emph{OQ}}(C,\lambda)}\\
&=  \min_{\frac{C}{m}<n\le\frac{N}{m}}{n^2m^2\inf_{\gamma>0}{\gamma\Lambda^*(\frac{C+1/\gamma}{nm},\lambda)}}.
\end{aligned}
\end{equation}
When $C=1$, the dominant mode is $(n^*m,\lambda,1,n^*mB)$ where $n^*\triangleq \arg\min_{C/m<n\le N/m}{n^2m^2\inf_{\gamma>0}{\gamma\Lambda^*(\frac{C+1/\gamma}{nm},\lambda)}}$.

The final result of Equation \ref{eq:mx1} looks very similar to that of Equation \ref{eq:cq}, just with fewer choices of $n$ during minimization. However, the higher start of the lowest valid overflow mode $n^*_{\min}m=m$ in \emph{PCQ-GLQF} rather than $n^*_{\min}=2$ in \emph{CQ-LQF} contributes to the effective increase in buffer size by a factor of $\frac{m}{2}$, especially when $\mu$ is small. $E_{32,(4\times 1)}^{\emph{PCQ-GLQF}} (1,\lambda)$ is plotted in Fig. \ref{fig:mx1}. The dominant mode drops to $n^*m=4$ when $\mu<0.752$ for \emph{PCQ-GLQF}, as opposed to $n^*=2$ when $\mu<0.8$ for \emph{CQ-LQF}. Also, $E_{32,(4\times 1)}^{\emph{PCQ-GLQF}}(1,\lambda)$ is much larger than $E_{32}^{\emph{CQ-LQF}}(1,\lambda)$ for low traffic load. Further improvements with larger $m$ values can be expected from Fig. \ref{fig:EnOQ}. 


Next, we turn to $1\times m$ pooling patterns. Following a similar approach as for \emph{CQ-LQF} in Section \ref{sec:inefficiency}, we consider OQ sub-systems. In addition to the number of crosspoints overflowing together ($n$), we also consider the number of outputs that are serving cells when overflow takes place ($m'$):
\begin{equation}
\label{eq:1xm}
\begin{aligned}
&E_{N,(1\times m)}^{\emph{PCQ-GLQF}} (C,\lambda)\\
&\triangleq \lim_{B\rightarrow \infty}{-\frac{1}{B}\log{P(\max_{1\le i \le N}{\sum_{j=1}^{m}{Q_{ij}}}>mB)}}\\
&=\min_{1\le m'\le m}{\lim_{B\rightarrow \infty}{-\frac{1}{B}\log{P(\max_{1\le i \le N}{\sum_{j=1}^{m'}{Q_{ij}}}>mB)}}}\\
&=\min_{1\le m'\le m}{m\lim_{B\rightarrow \infty}{-\frac{1}{B}\log{P(\max_{1\le i \le N}{\sum_{j=1}^{m'}{Q_{ij}}}>B)}}}\\
&=\min_{1\le m'\le m}{mE_{N}^{\emph{CQ-LQF}}(m'C,m'\lambda)}\\
&=\min_{1\le m'\le m}{\min_{m'C<n\le N}{mE^{\emph{OQ}}_{n}(m'C,m'\lambda)}}\\
&=\min_{1\le m'\le m}{\min_{m'C<n\le N}{n^2m\inf_{\gamma>0}{\gamma\Lambda^*(\frac{m'C+1/\gamma}{n},m'\lambda)}}}.
\end{aligned}
\end{equation}
The buffer overflow exponent $E_{32,(1\times 4)}^{\emph{PCQ-GLQF}} (1,\lambda)$ and its dominant mode $(n^*,m'^*\lambda,m'^*,n^*mB)$ are plotted in Fig. \ref{fig:1xm}. This is just a scaled version of Fig. \ref{fig:E32LQF}, in which all exponents are multiplied by $4$, while the turn point remains the same. It turns out that the dominant mode always has $m'^*=1$, which means only $1$ output is active upon overflow and pooling enlarges the buffer size by $4$ effectively. Meanwhile, $n^*=32$ when traffic load is high, and $n^*=2$ otherwise (this is the lowest possible overflow mode when $m'^*=1$).

Finally, based on Equations \ref{eq:mx1} and \ref{eq:1xm}, we derive the buffer overflow exponent for the generic $w\times r$ pooling pattern:
\begin{equation}
\label{eq:wxr}
\begin{aligned}
&E_{N,(w\times r)}^{\emph{PCQ-GLQF}} (C,\lambda)\\
&\triangleq \lim_{B\rightarrow \infty}{-\frac{1}{B}\log{P(\max_{1\le I \le \frac{N}{w}}{\sum_{i=(I-1)w+1}^{Iw}{\sum_{j=1}^{r}{Q_{ij}}}>wrB)}}}\\
&=\min_{1\le r'\le r}{\lim_{B\rightarrow \infty}{-\frac{1}{B}\log{P(\max_{1\le I \le \frac{N}{w}}{\sum_{i=(I-1)w+1}^{Iw}{\sum_{j=1}^{r'}{Q_{ij}}}>wrB)}}}}\\
&=\min_{1\le r'\le r}{r\lim_{B\rightarrow \infty}{-\frac{1}{B}\log{P(\max_{1\le I \le \frac{N}{w}}{\sum_{i=(I-1)w+1}^{Iw}{\sum_{j=1}^{r'}{Q_{ij}}}>wB)}}}}\\
&=\min_{1\le r'\le r}{rE_{N,(w\times 1)}^{\emph{PCQ-GLQF}}(r'C,r'\lambda)}\\
&=\min_{1\le r'\le r}{\min_{\frac{r'C}{w}<n\le \frac{N}{w}}{rE_{nw}^{\emph{OQ}}(r'C,r'\lambda)}}\\
&=\min_{1\le r'\le r}{\min_{\frac{r'C}{w}<n\le \frac{N}{w}}{rn^2w^2\inf_{\gamma>0}\gamma\Lambda^*(\frac{r'C+1/\gamma}{nw},r'\lambda)}}.
\end{aligned}
\end{equation}
$E_{32,(2\times 2)}^{\emph{PCQ-GLQF}} (1,\lambda)$ and the dominant $(n^*w,r'^*\lambda,r'^*C,n^*wrB)$ are plotted in Fig. \ref{fig:wxr}. This turns out to be another scaled version of Fig. \ref{fig:E32LQF}, in which all exponents are multiplied by $2$. The dominant $r'$ is still $1$, while $n^*w=32$ when traffic load is high, and $n^*w=2$ otherwise (this is the lowest possible overflow mode when $r'^*=1$).

%


\section{Scheduling Design \& Packet Ordering for Load Balancing and Deflection routing} 
\label{sec:CCQ}

In \cite{cqs,radonjic01}, it has been recognized that LQF provides a lower packet drop rate for the basic CQ switch than any other simple scheduling algorithms like random, RR and OCF. However, its performance can still be far worse than an OQ switch with the same total buffer space, if the incoming traffic is bursty or non-uniform. In this section, we first propose a scheme that allows different crosspoints in the same daisy chain to share packets evenly using load balancing and deflection routing, then we apply OCF and RR-based scheduling algorithms to ensure correct packet ordering.

\subsection{Oldest-Cell-First Scheduling in a CCQ switch}
\label{sec:OCF}

OCF is a popular scheduling algorithm which always picks the oldest cell to serve. Compared with LQF, OCF usually incurs a larger packet drop rate since it does not always serve the buffer that is most likely to overflow. Compared with RR, OCF has a much more complex implementation since it requires repeated comparisons of time-stamps at each time slot. Despite these disadvantages, OCF is still attractive since it can easily maintain the packet order across all flows. This advantage makes OCF a good candidate to solve the mis-sequencing problem caused by load balancing and deflection routing. The performance loss due to using OCF rather than LQF can be negligible since load balancing and deflection routing already do a good job in equalizing the buffer utilizations.

In this scheme, we use the two-stage CCQ switch. Every incoming cell is assigned a time-stamp to record its arrival time. Each crosspoint needs to maintain the buffered cells in the order of non-decreasing time-stamps (i.e., first-come-first-serve). Then the output schedulers will only need to compare the time-stamps of HOL cells to determine the oldest one in each time slot.

The detailed scheme for \emph{CCQ-OCF} is described below:
\begin{itemize}
\item
\textbf{\emph{Arrival Phase:}} 
At time $t$, for each input $i$, if there is a newly arriving cell destined to output $j$, then after passing the load-balancing stage that connects input port $i$ to intermediate port $i+t$, it is directly sent to crosspoint $(i+t,j)$ of the second stage. If the buffer is not full, i.e., $b_{i+t,j}<B$, the new cell is accepted and buffered at the TOL with time-stamp $t$. Otherwise, this overflowing cell is dropped.

\item
\textbf{\emph{Departure Phase:}} 
For each output $j$, if there is at least one non-empty crosspoint buffer $(*,j)$, the output scheduler picks the one with the oldest HOL cell, and serves this cell.

\item
\textbf{\emph{Deflection Phase:}} 
Each crosspoint $(i,j)$ does the following step by step:
1) Report buffer occupancy $b_{ij}$ to its successor crosspoint $(i+1,j)$;
2) Receive a buffer occupancy report $b_{i-1,j}$ from its predecessor crosspoint $(i-1,j)$;
3) If $b_{ij}>b_{i-1,j}$, deflect the HOL cell to its predecessor crosspoint $(i-1,j)$;
4) Receive a deflected cell from its successor crosspoint $(i+1,j)$. If there is one, insert the deflected cell into the ordered queue according to its time-stamp, which could be completed within $O(\log{B})$ time using a self-balancing tree.

\end{itemize}

\subsection{Round-Robin Scheduling with Counter Alignment}
\label{sec:RR}

In the previous section, the OCF scheduling algorithm has been used to maintain the correct packet order. This method is straightforward and promising, but requires considerable computation due to repeated sorting in each time slot. On the other hand, the global packet ordering guaranteed by OCF is too strict, since we only need per-flow packet ordering. In this section, we propose a new scheme that relies on a less-demanding RR polling algorithm and an explicit notification mechanism between adjacent crosspoints to preserve per-flow packet ordering. The underlying idea is partly inspired by the Mailbox Switch \cite{mailbox} and Padded Frame \cite{frame02}, but it is implemented in a very different way here that avoids extra delays.

\subsubsection{Wait-Counter and RR-Counter}
\label{wait}

In this scheme, every crosspoint maintains a ``wait-counter'' for each of its buffered cells, denoted by $W_{ij}(k)$, in which $1\le k \le b_{ij}$ is the position of that cell. Another anticipatory wait-counter for the next incoming cell, denoted by $W_{ij}(b_{ij}+1)$, is also maintained by crosspoint $(i,j)$. When a new cell arrives at $(i,j)$, it is assigned $W_{ij}(b_{ij}+1)$ upon acceptance. Then $b_{ij}$ gets incremented, and a new anticipatory wait-counter is generated as $W_{ij}(b_{ij}+1)\leftarrow W_{ij}(b_{ij})+1$. Wait-counters $W_{ij}(k)$ are ever-increasing with $k$, but the carries may be dropped when they exceed a sufficiently large value to solve the grow-to-infinity problem.

As a counterpart of the wait-counters, we also let each output $j$ maintain a ``RR-counter'' $R_j$, in addition to its arbiter position $1\le A_j \le N$ which always points to the last crosspoint it has polled. $R_j$ tracks the number of RR pooling cycles that arbiter $j$ has performed, and is incremented during each cycle when $A_j=1$. $R_j$ also grows to infinity and must be treated in the same way as $W_{ij}(k)$.

The RR-counters and wait-counters $W_{ij}(k)$ are always maintained in non-decreasing order, so that $R_j \le W_{ij}(k)\le W_{ij}(k+1)$ for any $1\le i,j \le N$ and $1 \le k \le b_{ij}$ at any time. $R_j$ would not exceed $W_{ij}(1)$ when $b_{ij}>0$, otherwise $W_{ij}(1)$ is set to $R_j+1$ each time it is polled by the output. 

An arbitrary cell $k$ stored at a non-empty crosspoint $(i,j)$ is eligible to leave the switch, if and only if $W_{ij}(k)=R_j$, thus crosspoint $(i,j)$ must refrain from being served by output $j$ until its HOL cell becomes eligible. 

\subsubsection{Counter-Alignment Notification}
\label{alignment}

We also design an explicit counter-alignment notification mechanism, which coordinates the correct packet ordering under load balancing. Such a notification is initiated by any crosspoint $(i,j)$ upon acceptance of a newly arriving cell. It is then passed down to $(i+1,j)$ and subsequent crosspoints along the daisy chain. Upon reception, the receiver crosspoint examines the contents, make necessary updates to its own anticipatory wait-counter, and determine whether to drop the notification or to relay it to subsequent crosspoints. 

Information contained in a notification message consists of two parts: a counter-alignment field $CA_{ij}$, which indicates the minimum wait-counter for the next incoming cell to crosspoint $(i+1,j)$, and a source-of-notification field $SN_{ij}$, which denotes the crosspoint that has initiated the message.

Specifically, when crosspoint $(i,j)$ accepts a new cell, it immediately initiates a counter-alignment notification with $CA_{ij} \leftarrow W_{ij}(b_{ij})$ (increment if $i=N$) and $SN_{ij}=i$, and sends it to the successor crosspoint $(i+1,j)$ in the same daisy chain.

Then for crosspoint $(i+1,j)$, if $CA_{ij}\ge W_{i+1,j}(b_{i+1,j}+1)$ and not $SN_{ij}=i+1$ (discard the message if it has traversed the daisy chain and come back to its origination), it updates $W_{i+1,j}(b_{i+1,j}+1)\leftarrow CA_{ij}$, and decides to relay the notification message with $CA_{i+1,j}\leftarrow CA_{ij}$ (increment if $i+1=N$) and $SN_{i+1,j}\leftarrow SN_{ij}$ to its own successor $(i+2,j)$ in the next time slot, if by that time it has not accepted a new cell and generated a new notification message.

In this way, the mis-sequencing problem caused by load balancing can be solved. Cells of the the same flow are always assigned with non-decreasing wait-counters through just-in-time notifications between consecutive arrivals.

\subsubsection{Deflection Routing with Counter Preserved}
\label{deflect}

Deflection routing may also introduce mis-sequencing. With wait-counters, it is straightforward to resolve the issue.

Similar to \emph{CCQ-OCF}, each crosspoint $(i,j)$ is allowed to deflect one HOL cell to its predecessor $(i-1,j)$ in each time slot if $b_{ij}>b_{i-1,j}$, except that crosspoint $(A_j,j)$ does not deflect if $W(A_j,j,1)=R_j$ (which means its HOL cell is already eligible to leave). The deflected cell carries its own wait-counter $DW_{ij} \leftarrow W_{ij}(1)$ (decrement if $i=1$) with it. When crosspoint $(i-1,j)$ receives the deflected cell, it compares $DW_{ij}$ with its own cells, and inserts the deflected cell to the appropriate position to maintain non-decreasing order of wait-counters. If it has one or more cells with wait-counters equal to $DW_{ij}$, the deflected cell is inserted behind all of them to preserve their relative order of departure. In case $DW_{ij}\ge W_{i-1,j}(b_{i-1,j}+1)$, update $W_{i-1,j}(b_{i-1,j}+1)\leftarrow DW_{ij}+1$.

Now that there may be multiple cells with the same wait-counters at each crosspoint $(i,j)$, output $j$ must adopt a batch RR algorithm, serving all cells $k$ at crosspoint $(i,j)$ with $W_{ij}(k)=R_j$ before proceeding to the next eligible crosspoint. In this way, deflection routing will not alter the order of cells to be served.

\subsubsection{CCQ-RR Scheme}
\begin{itemize}
\item \textbf{\emph{Arrival Phase:}} 
Same as in \emph{CCQ-OCF} except that the wait-counters are assigned and updated according to Section \ref{wait} instead of the time-stamps.

\item \textbf{\emph{Notification Phase:}} 
Each crosspoint $(i,j)$ sends and receives a counter-alignment notification message according to Section \ref{alignment}.

\item \textbf{\emph{Departure Phase:}} 
Each output $j$ polls its associated crosspoints $(*,j)$ in an exhaustive RR fashion, starting from its final position $A_j$ in the previous time slot. The polling process continues until output $j$ serves an eligible crosspoint with $W_{ij}(1)=R_j$, or it finds all buffers empty.

\item \textbf{\emph{Deflection Phase:}} 
Same as in \emph{CCQ-OCF} except that wait-counters take the place of time-stamps according to Section \ref{deflect}.
\end{itemize}

An example is illustrated in Fig. \ref{fig:exampleIII}. Different flows are marked with different colors and alphabets, e.g., $yellow-a$. The time-stamps (for illustration, not required in implementation) are indicated by integer subscripts, e.g., 1,2,3. Wait-counters are represented by their positions on the time-line, while vacancies (cross-marked squares) in the time-lines do not occupy real buffer positions.  During the arrival phase at $t=1$, newly arriving cells $a_2$ and $c_2$ are tagged with wait-counters $W_{2,j}(1)=0$ and $W_{4,j}(2)=1$ respectively. Next, during the notification phase, crosspoint $(2,j)$ initiates a counter-alignment notification $CA_{2,j}\leftarrow W_{2,j}(1)=0$ for the newly accepted cell $a_2$ and sends it to successor $(3,j)$, but this notification is discarded because $CA_{2,j}=0<W_{3,j}(2)=1$. On the other hand, crosspoints $(4,j)$ also initiates a counter-alignment notification $CA_{4,j}\leftarrow W_{4,j}(2)+1=2$ (note that $i=4=N$ here), and crosspoint $(1,j)$ accepts it, updates $W_{1,j}(2)\leftarrow CA_{4,j}=2$ (a vacancy is created in the time-line), and decides to relay it in the next time slot. Then during the departure phase, the first eligible cell $b_1$ with $W_{1,j}(1)=0=R_j$ is served by the output, pushing all subsequent cells ahead. Finally, during the deflection phase, crosspoint $(1,j)$ receives the HOL cell $a_2$ from successor $(2,j)$ with $DW_{2,j}\leftarrow W_{2,j}(1)=0$ and inserts it to the HOL with $W_{1,j}(1)\leftarrow DW_{2,j}=0$, while crosspoint $(3,j)$ receives the HOL cell $d_1$ from crosspoint $(4,j)$ with $DW_{4,j}\leftarrow W_{4,j}(1)=0$ and places it behind cell $c_1$ with the same wait-counters $W_{3,j}(1)=W_{3,j}(2)\leftarrow DW_{4,j}=0$ (two cells occupy a single slot in time-line). As a result, the newly arriving cells $a_3$ and $c_3$ to arrive at $t=2$ will be assigned with wait-counters $W_{3,j}(3)\leftarrow W_{3,j}(2)+1 =1$ and $W_{1,j}(1)=2$ respectively. As we can see, cell order is maintained by just-in-time notifications and intentional vacancies, so that the wait-counters assigned to cells of the same flows are always non-decreasing. The cells will leave the switch in the order of $b_1$, $a_2$, $c_1$, $d_1$, $a_3$, $c_2$, $c_3$, assuming no more new cells.

%
%
%

\begin{figure}[ht]
\begin{minipage}[t]{3.2 in}
\centering \subfigure[Initial case at time $t=1$.]{
\includegraphics[width=3.2 in]{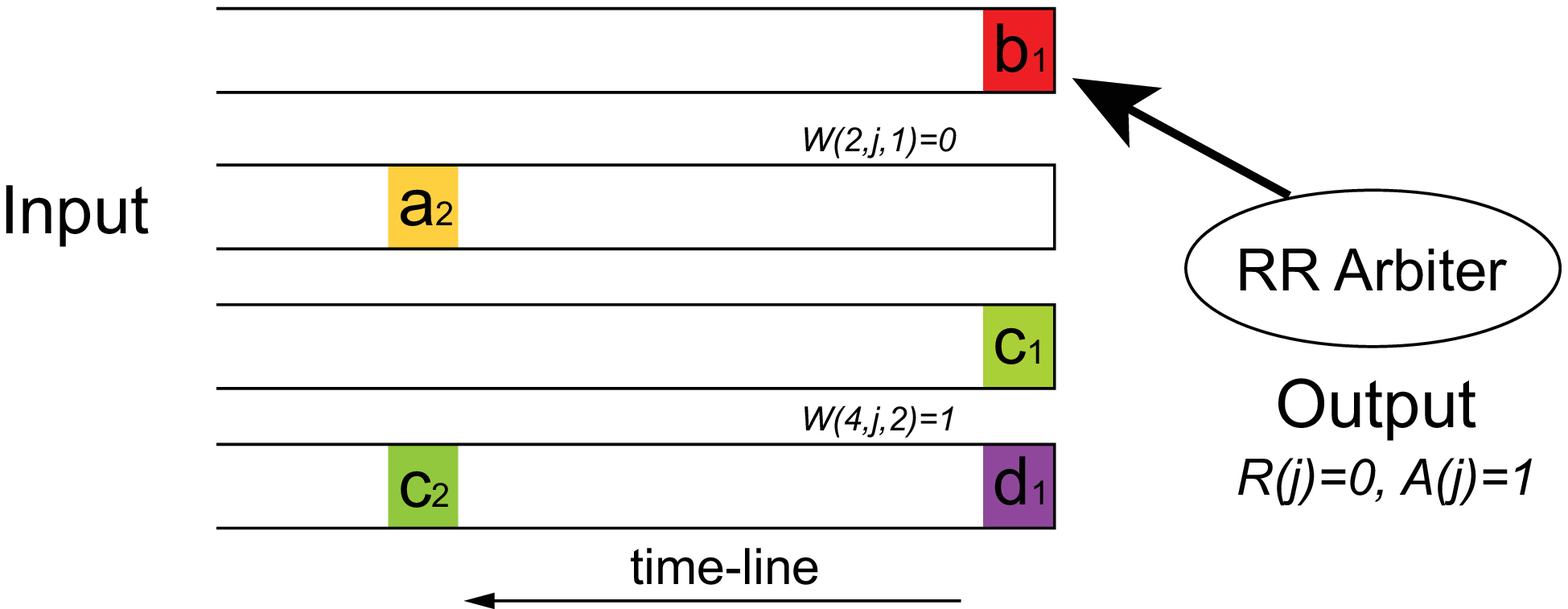}
\label{fig:IIIA}}
\end{minipage}
\begin{minipage}[t]{3.2 in}
\centering \subfigure[Changes until time $t=2$.]{
\includegraphics[width=3.2 in]{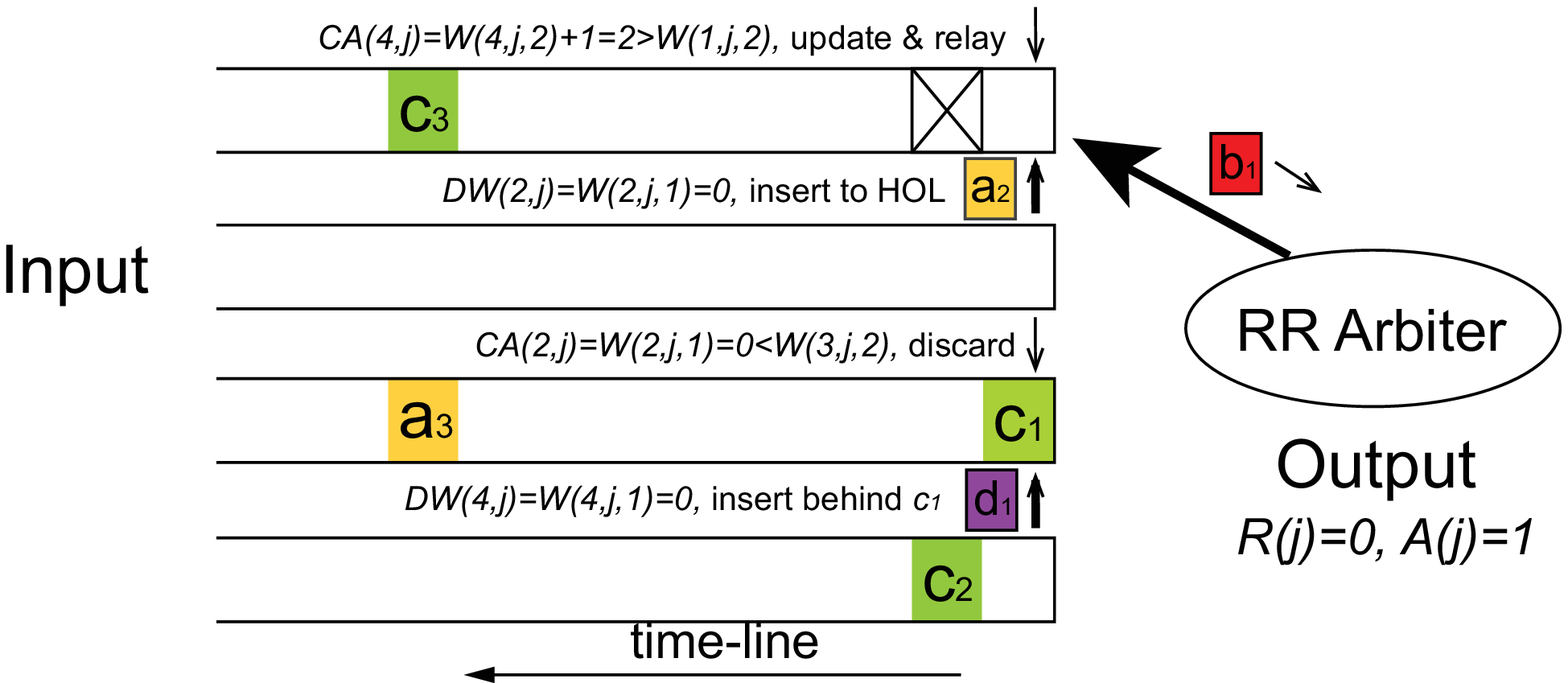}
\label{fig:IIIDD}}
\end{minipage}
\caption{An example of \emph{CCQ-RR}.}
\label{fig:exampleIII}
\end{figure}

\subsection{Properties of CCQ-RR}
\label{sec:feature}

\textbf{\emph{Property 1:}} The proposed \emph{CCQ-RR} scheme is work-conserving, if the maximum number of deflections is restricted to $K$ and each output can perform $N+K+1$ polls in each time slot. Interestingly, through our lengthy simulations, none of the cells is deflected more than $K=N-1$ times.

First, consider the situation without deflection routing. Pick any arbitrary cell $\mathcal{X}$ that arrives at crosspoint $(i,j)$ and gets wait-counter $W_{ij}(k)$.
\begin{itemize}
\item
If $W_{ij}(k)$ was updated upon acceptance of a newly arriving or deflected cell $\mathcal{Y}$, then $\mathcal{Y}$ must have been exactly $N+1$ polls away at that time. 

\item 
If $W_{ij}(k)$ was updated through counter-alignment initiated for cell $\mathcal{Y}$, then $\mathcal{Y}$ must have been at most $N$ polls away at that time, since otherwise the counter-alignment notification should have already been discarded after traversing the daisy chain. 

\item
Otherwise, $W_{ij}(k)$ must have been updated when crosspoint $(i,j)$ was empty through $W_{ij}(1)=R_j+1$, then $k=1$ and it is at most $N$ polls away from the output arbiter.
\end{itemize}

Summing up all three conditions, the output arbiter needs at most $N+1$ polls (starting from its last polled crosspoint) in each time slot to ensure it is work-conserving.

We next take deflection routing into consideration. In fact, since the direction of deflection routing reverses the RR polling order, cells are always pushed closer to the arbiters, while the gaps between two consecutive cells (in the order of departure) are enlarged by at most $K$. As a result, each output arbiter needs at most $N+1+K$ polls to ensure work-conserving.

\textbf{\emph{Property 2:}} Cells of the same flow always leave the switch in the same order as they arrive.

For load balancing, cell order is preserved through just-in-time counter-alignment notifications between any two consecutive arrivals of the same flow. In terms of deflection routing, it will not alter the order of departure if the wait-counters are preserved and adjusted when necessary. These are elaborated in Sections \ref{alignment} and \ref{deflect}, as long as some boundary conditions are taken care of. Specifically, the last crosspoint $(N,j)$ in each daisy chain $j$ must always increment the counter-alignment field $CA(N,j)$, whereas crosspoints $(1,j)$ must always decrement the wait-counter of its deflected cell $DW(1,j)$, so as to match with the starting point of a new RR polling cycle.

%
%

\textbf{\emph{Property 3:}} The worst-case time complexity at each crosspoint is $O(\log{B})$ in each time slot, and each output scheduler can find the next eligible HOL cell in $O(\log{N})$ time.

As mentioned before, the crosspoints need to maintain the cells in non-decreasing order of wait-counters. Since such ordering may only be disturbed upon cell arrival, departure and deflection, we have:
\begin{itemize}
\item
Newly incoming cells are always be placed at the tail of line and are assigned the largest wait-counters so far at this crosspoint. Thus cell arrival does not break the ordering and there is no need for comparisons here.

\item
Only HOL cells can be served. These cells always have the smallest wait-counters. Thus cell departure does not break the ordering either.

\item
Only the oldest cells (HOL cells), can be deflected from highly-utilized crosspoints (sender) to their predecessor crosspoints (receiver). They always have the smallest wait-counters at the senders. Thus deflection does not break the ordering at these senders.

\item
In each time slot, each crosspoint may receive at most one deflected cell from its successor. This deflected cell is then searched and inserted into the pre-ordered queue at the receiver according to its wait-counter.

\end{itemize}

Taking the arrival and departure phases into account, each crosspoint needs to perform $O(1)$ search, insertion, and deletion operations in each time slot. Besides, $O(1)$ additional updates to the anticipatory wait-counters need to performed. All these can be accomplished in $O(\log{B})$ time using a self-balancing binary search tree.

In terms of the output scheduler, each RR arbiter may find the next eligible crosspoint within $O(\log{N})$ time using a hardware-based priority encoder \cite{arbiter} (typically a few nanoseconds). Although the magnitude of time complexity for RR appears to be the same as that for OCF, the constant factor can be much smaller, and it is widely recognized that RR is much easier to implement than OCF. On the other hand, in order to utilize the priority encoder, each output arbiter $j$ may need to broadcast its RR-counter $R_j$ and arbiter position $A_j$, so that each crosspoint may determine its own eligibility in a distributed manner.

\textbf{\emph{Property 4:}} The maximum span of wait-counters that coexist in a single daisy chain is $NB+\lceil K/N\rceil$, if the maximum number of deflections is restricted to $K$. Therefore, the overhead of wait-counters can be bounded by $\log_2{(NB+\lceil K/N \rceil)}\approx 16 bits$ for $N=128$, $B=455$, and $K=N-1$.

First we assume that the wait-counters are ever-increasing. Then we notice that the largest-ever wait-counter can only be generated by new arrivals but not deflections, departures, or notifications. To be more specific, each incoming cell increases the largest-ever wait-counter by either $1$ or $0$. Therefore, without deflection routing, the difference between the largest and the smallest wait-counters that coexist in the system is bounded by the maximum number of cells $NB$. Going one step further by taking deflection routing into account, the smallest wait-count may decrease by at most $\lceil K/N \rceil$ if the number of deflections is bounded by $K$, thus the span of wait-counters that coexist in a single daisy chain is bounded by $NB+\lceil K/N\rceil$.

In practical implementation, we can use binary values to represent these different wait-counters. Since the wait-counters are compared with the RR-counters to determine eligibility, carries can be dropped if the number of bits is already sufficient to avoid overlaps and confusions, i.e., $NB+\lceil K/N\rceil<2^{\emph{overhead}}$.


\subsection{Interworking of Load Balancing and Deflection Routing}

Till now we have successfully enabled both load balancing and deflection routing on the augmented CCQ switching architecture, and designed scheduling algorithms to cope with both functionalities. Then we discuss the feasibility of further exploiting the interworking between load balancing and deflection routing by disrupting the order consistencies, as motivated in Section \ref{sec:deflection}. Assuming fixed RR polling order, we change the load balancing or deflection routing order.

1) \textbf{Changing load balancing order:} Under the OCF policy, load balancing can follow any order freely, either deterministic or random (using a random order also helps fighting adversarial traffic patterns), as long as the traffic distribution is uniform. Under the RR scheduling with counter alignment, counter notifications must be sent prior to future cell arrivals according to the load balancing order. Since no instantaneous communications should be required between the inputs and the switching fabric, the load balancing order must either be deterministic or pseudo-random based on a common generator and seed shared by all inputs and the switching fabric. In addition, when sending out or forwarding notifications, the sender must increment $CA_{ij}$ whenever the receiver has a smaller index. The new load balancing order requires a new logical notification path among the crosspoints, but can be mapped onto existing physical connections in the crossbar;

2) \textbf{Changing deflection routing order:} This is feasible under the OCF policy. Since the service order will not be disturbed anyway, deflection can appear in any form as long as the speedup constraints are met. For RR scheduling, uni-laterally changing the deflection routing order may disturb the correct cell ordering and thus is infeasible.

\section{Scheduling Design \& Contention Resolution for Buffer Pooling}
\label{sec:pooling}

In addition to load balancing and deflection routing, buffer pooling may also help mitigate the buffer space limitation. By aggregating crosspoints together, statistical multiplexing gains could be achieved across different inputs and outputs. However, buffer pooling also introduces some new challenges, and how to design pooling patterns and scheduling algorithms for the PCQ switch remains a problem to be solved.

\subsection{Pooling Patterns}

The pooling pattern has a large impact on the performance. In Equation \ref{eq:wxr} of Section \ref{sec:pooling}, we have already established an expression for the buffer overflow exponent $E_{N,(w\times r)}^{\emph{PCQ-GLQF}} (1,\lambda)$ of a generic $w\times r$-pooled CQ switch, and the dominant overflow mode is $(n^*w,r'^*\lambda,r'^*C,n^*wrB)$.

\begin{enumerate}
\item Under high traffic load, all (pooled) crosspoint buffers associated with the same output tend to overflow at the same time, while different outputs tend to overflow at different times. Therefore, the dominant mode is always $(N,\lambda,1,NrB)$, thus a larger $r$ corresponds with a better buffer sharing effect and a lower overflow probability;

\item Under low traffic load, it is more likely that the (pooled) crosspoint buffers will overflow separately at the lowest possible mode, and the dominant mode would be $(\max\{2,w\},\lambda,1,\max\{2,w\}rB)$. In this case, both $w$ and $r$ contribute to buffer sharing. However, a larger $w$ also means more arrival processes with rate $\lambda$, which results in a higher traffic load. Consequently, increasing $r$ is still more effective than increasing $w$.
\end{enumerate}

The intuition of these results can be attributed to the fact that LQF has already balanced the buffer utilizations within the daisy chains, while buffer pooling can not only improve buffer sharing among the balanced crosspoint buffers within the daisy chains, but also extend the buffer sharing effect across multiple daisy chains. The effects of LQF and buffer pooling can be complimentary, and thus it is more effective to pool buffers associated with different outputs.



The above results are derived without considering the speedup and complexity requirements. In practice, a $w\times r$ pooling pattern would have $w$ input contentions and $r$ output contentions. If we only use memory speedup to resolve the contentions, and the two kinds of speedup are equally demanding, then there would be a total speedup requirement of $s_w+s_r=w+r$. 
In case that the pooling gain is dominated by the pooling size $w\times r=m$, it would be most efficient if $w=r$. 
This is a simplified analysis which overlooks the difference between input and output contention. In fact, input contention is less flexible and demands more hardware speedup, whereas output contentions can be solved with both higher hardware speedup and more sophisticated scheduling. So there is a tradeoff between hardware complexity and software complexity. Also, the marginal benefit of allocating more memory-read speedup $s_r$ decreases dramatically after it passes a threshold $s^*_r=s_w$, because any larger value of $s_r$ would be more than what is required to keep the queues stable and provides little help in further reducing the cell drop rate.

Summing up, pooling buffers across different outputs is always more effective in avoiding overflow, and less demanding in hardware speedup, but requires more sophisticated software scheduling. On the other hand, buffer pooling within the daisy chains can still be useful when hardware speedup is affordable.


\subsection{Contention Resolution}
\label{sec:contention}
As we have mentioned before, there may be both input and output contentions after buffer pooling. For a $w\times r$ pooling pattern, as many as $w$ inputs and $r$ outputs may request memory-write/read at the same time. One straightforward way to accomodate such simultaneous memory accesses is to implement sufficient hardware speedup, which could be as high as $w+r$. However, this approach is neither practical nor efficient. In high-speed Internet core switches, the line rate of each single input/output already operates close to the hardware limits. On the other hand,  over-provisioning for the worst case provides only marginal gains.


For input contention, extra speedup is the only solution to avoid memory blocking and packet drops. However, a full memory-write speedup of $s_w=w$ might be wasteful. In fact, the probability that $k$ out of $m$ crosspoints in a common pool receive cells at the same time decays exponentially with $k$. Therefore, full speedup is not always necessary.

For output contention, there is more scope for innovation. For $w\times r$ buffer pooling, as many as $r$ outputs may concurrently try to serve different cells buffered in the same pool under LQF/RR/OCF policies. However, we notice that these cells are just their first-choices, and such choices are subject to compromise as long as better performance can be achieved. To be specific, we may consolidate $r$ outputs (connected to the same pool) into a group, and perform joint scheduling for their departure processes. Denote by $\{I,J\}$, $1\le I\le \frac{N}{w}$, $1\le J \le \frac{N}{r}$, the buffer pool that aggregates crosspoints $(i,j)$ with $i=(I-1)w+1,...,Iw$ and $j=(J-1)r+1,...,Jr$, with a total buffer size $p(I,J)\le P\triangleq wrB$. Assume each output $j$ has a \emph{preferred list} of buffer pools that it would like to serve, and each buffer pool $\{I,\lceil\frac{j}{r}\rceil\}$ carries a positive weight $W_{I,j}$ if it is in the preferred list of output $j$, and zero weight otherwise. The weight can be a function the queue length under the LQF rule, or a function of the order of departures under OCF or RR policies, etc. Then the output contention resolution problems can be formulated as a variation of the well-known MWM problem: at each time slot $t$, given a $w\times r$ pooling pattern and a $\frac{N}{w}\times N$ weight matrix $W_{I,j}(t)$, find a match $S$ between output $j$ and buffer pool $\{I,\lceil\frac{j}{r}\rceil\}$ under the constraint of memory-read speedup, 
so that the total weight $\sum_{(I,j)\in S}{W_{I,j}(t)}$ is maximized.
\begin{gather}
\max_{S}{\sum_{(I,j)\in S}{W_{I,j}(t)}}\\
\text{s.t.}~\sum_{j=(J-1)r+1}^{Jr}{\textbf{1}_{_{(I,j)\in S}}}\le s_r,~ 
\text{for}~I=1,...,\frac{N}{w},~J=1,...,\frac{N}{r}\\
\text{and}~\sum_{I=1}^{N/w}{\textbf{1}_{_{(I,j)\in S}}}\le 1,~
\text{for}~j=1,2,...,N
\end{gather}

The MWM problem has been well studied, and there exist a wide variaty of optimal or heuristic solutions in literature, so we do not repeat them here. One might question whether solving a MWM problem in a centralized way at each time-slot would be too costly. Here we argue that it could much less demanding for these reasons:
1) The system-wide MWM problem of size $N$ can be divided into sub-problems of size $r$, because only every $r$ outputs that are connected to the same buffer pools need to be jointly scheduled;
2) Batch scheduling \cite{batch}, iterative algorithms \cite{islip}, etc., can be applied in solving the MWM problem to reduce the computation complexity at each time-slot;
3) Maximal matching \cite{islip}, randomized matching \cite{disquo}, or other heuristic algorithms may also provide near-optimal performances, but at a much lower cost.

In the next section, a GLQF scheduling scheme with contention resolution for the PCQ switch will be proposed. The proposed OCF and RR algorithms that support load balancing and deflection routing may also be integrated into PCQ switches. However, separate virtual input queues need to be maintained, and additional re-sequencing buffers may be needed at the outputs due to contention resolution.

\subsection{Generalized Longest-Queue-First Scheduling with Contention Resolution by Maximum-Weight-Matching}


In Section \ref{sec:wxr}, we assumed an ideal GLQF policy for PCQ switches, which always serves the longest queues without taking the speedup limit into account. Here we propose a practical GLQF scheduling with contention resolution by MWM, \emph{PCQ-GLQF-MWM}, for each output $j$ to decide which pooled buffer $\{I,\lceil\frac{j}{r}\rceil\}$ to serve.
\begin{itemize}
\item
\textbf{\emph{Arrival:}} For each input $i$, if there is a newly arriving cell destined to output $j$, it is directly sent to crosspoint $(i,j)$ which resides in buffer pool $\{\lceil\frac{i}{w}\rceil,\lceil\frac{j}{r}\rceil\}$. If buffer pool is not full, the new cell is accepted and buffered at the tail of line (TOL). Otherwise, a cell is dropped according to additional buffer management rules.

\item
\textbf{\emph{Departure:}} 
1) Each output $j$ sorts all connected buffer pools $\{I,\lceil\frac{j}{r}\rceil\}$, $I=1,2,...,\frac{N}{w}$, in non-increasing order of the number of cells destined to it, i.e., $\sum_{i=(I-1)w+1}^{Iw}b_{ij}$, and picks the first $\max(\frac{N}{w},r)$ pools into its preferred list;
2) Solve the MWM problems for each group of $r$ outputs under speedup constraint of $s_r$ as in Section \ref{sec:contention};
3) Each output $j$ serves a buffer pool according to the optimal matching solution $S$ derived in the previous step. The specific cell to be served is determined by the additional buffer management rules.
\end{itemize}

%

\section{A Comprehensive Buffer Sharing Solution for Crosspoint-Queued Switches}
\label{sec:framework}

Till now we have described all building blocks for efficient buffering and scheduling in a CQ switch. The basic CQ switching architecture is introduced in Section \ref{sec:basic}. Three different buffer sharing techniques -- load balancing, deflection routing, and buffer pooling, as well as the augmented switching architectures -- CCQ and PCQ structures, are proposed in Section \ref{sec:modified}. Their effects in combating unbalanced buffer utilizations are also analyzed using the theory of large deviations and Markov model. In Sections \ref{sec:CCQ} and \ref{sec:pooling}, several practical scheduling schemes based on the legacy OCF, RR, and LQF policies are specially tailored for these buffer sharing techniques. The main takeaways as follows:
\begin{itemize}
\item The basic LQF policy always serves the longest queue that is most likely to overflow. It works well when buffer size is large, but is inefficient when there is limited space and the incoming traffic is bursty and non-uniform. The balancing effect of LQF is limited within a single output;
\item Load balancing re-distributes incoming traffic uniformly. It can transform non-uniform traffic into uniform traffic, and reduce the traffic burstiness at the same time. The out-of-sequence problem can be either solved by adopting OCF scheduling that incurs high cell overhead and comparison complexity, or by employing the proposed RR policy with a lower scheduling complexity but some modest architecture modification (mapping new logical connections to physical links) and an extra counter notification mechanism. Load balancing distributes incoming traffic within a single daisy chain;
\item Deflection routing is capable of re-arranging the cells after arrival, and has the potential of perfectly equalizing the buffer utilizations. Its effect is regional, and requires some time to propagate through the daisy chain. An extra memory-read and memory-write speedup is required at each crosspoint buffer, and new logical connections needs to be mapped. The out-of-sequence problem can be solved by the time-stamps in OCF or the wait-counters in the proposed RR policy. Deflection routing moves cells around within a single output;
\item Buffer pooling allows for dynamic sharing and allocation of buffers among crosspoints that are pooled together. For maximum performance improvement, it is recommended that buffer pooling should be done across multiple outputs rather than multiple inputs within one single output. Regarding hardware requirements, a $w\times r$ PCQ switch should have a memory-write speedup of $s_w=\min\{w,\frac{\log{P_{\emph{drop}}}}{\log{(wr/N)}}\}$ (where $P_{\emph{drop}}$ is the desired cell drop rate) and a memory-read speedup of $\min\{s_w,r\}$ to ensure low cell drop rates. In case $r>w$, the remaining $r-w$ output contentions can be resolved by either extra speedup or MWM scheduling. The buffer sharing effect can cross multiple outputs.
\end{itemize}

Then, we put forward a comprehensive buffer sharing solution for CQ switches in various cases:
\begin{enumerate}
\item If the incoming traffic is non-uniform or highly bursty, or if the switch size is very large, load balancing offers more improvement than deflection routing by re-distributing traffic evenly to a whole daisy chain. Otherwise, deflection routing or the LQF policy may better balance the utilizations within a single output;
\item If the buffer size is extremely limited, buffer pooling is the only way to improve performance. As the buffer size increases, load balancing starts gaining benefits from the law of large numbers, and deflection routing gets enough time to propagate through the daisy chain;
\item If scheduling complexity and memory speedup are restricted, the proposed RR algorithm with load balancing and deflection routing offers a simple but effective solution. Otherwise, the OCF policy is the most straightforward way to ensure correct packet ordering, and buffer pooling with contention resolution by MWM could further suppress the cell drop rate.
\end{enumerate}

\section{Discussion on Delay, Multicast and QoS}
\label{sec:discussion}

\subsection {Delay Performance}

Till now we have been focusing on the cell drop rate and the buffer utilization, and have successfully improved them through architecture and scheduling design. We now briefly discuss delay performance.

The (average) delay experienced by a cell in any work-conserving CQ switch is no higher than in an OQ switch with sufficient speedup under the same conditions (system size $N$, $B$, and arrival processes $\{X_t\}$), as long as this cell is accepted and delivered in both systems. We elaborate this result with the following sample path analysis for a single output:

1) If the buffer size is infinite, there will be no cell drops in either the CQ switch or the OQ switch. In the special case when both switches adopt the same OCF policy, every cell faces exactly the same queueing process and thus the same delay in both switches. More generally, as long as the schedulers are both work-conserving, the specific service order among different cells in the system only affects the delay distribution, but not the sum of delay experienced by all cells, and hence the average delay for each cell;

2) When the buffer size is finite, there will be more cell drops with the CQ switch than with the OQ switch. This can be further divided into various dropping policies, and we shall pick two typical ones -- HOL dropping and TOL dropping. The commonly-used TOL dropping policy simply rejects new cells when the buffer is full, and all accepted cells are always delivered. On the other hand, the HOL dropping policy accepts all cells, but drops HOL cells when the buffer is full. The dropped cells are never delivered, though they have already waited in the queue for some time. Such a drop-from-front strategy was proposed for TCP enhancements \cite{tcphol}.

First, it can be proved that the total buffer occupancy in the CQ switch can never exceed that in the OQ switch:
\begin{itemize}
\item Initially at $t=0$, $\sum_{i=1}^{N}{b_0^{\emph{CQ}}(i,j)}=b_0^{\emph{OQ}}=0$;

\item If $\sum_{i=1}^{N}{b_t^{\emph{CQ}}(i,j)}\le b_t^{\emph{OQ}}$ holds immediately before cell arrivals at time-slot $t$, then $\sum_{i=1}^{N}{b_{t+1}^{\emph{CQ}}(i,j)}\le b_{t+1}^{\emph{OQ}}$ is also valid at time $t+1$ according to the following two cases: 

a) If no cell is dropped by \emph{OQ}, then $b_{t+1}^{\emph{OQ}} = [b_t^{\emph{OQ}}+\sum_{i=1}^{N}{X_t(i,j)}-1]^+\ge [\sum_{i=1}^{N}{b_t^{\emph{CQ}}(i,j)}+\sum_{i=1}^{N}{X_t(i,j)}-1]^+\ge \sum_{i=1}^{N}{b_{t+1}^{\emph{CQ}}(i,j)}$; 

b) If at least one cell is dropped by \emph{OQ}, then $b_{t+1}^{\emph{OQ}}=NB-1\ge \sum_{i=1}^{N}{b_{t+1}^{\emph{CQ}}(i,j)}$.
\item Therefore, whenever a cell arrives, it sees an equally or less occupied CQ switch than OQ switch.
\end{itemize}
 
For TOL dropping and OCF scheduling, the delay experienced by a certain accepted (and also delivered for sure) cell is simply the number of cells that are already in the switch when this cell arrives at the switch, so its delay in the CQ switch never exceeds that in the OQ switch. Following the same argument as in 1), as long as the scheduling policy is work conserving, the average delay experienced by each cell in the CQ switch is no higher than in the OQ switch. 

For HOL dropping, things become more complicated because an accepted cell may still be dropped and not delivered. Under OCF scheduling, the delay experienced by an arbitrary accepted and delivered cell is not only determined by the queue size upon arrival, but also affected by the number of cells that are dropped after its arrival and before its departure.  Suppose that the CQ (or OQ) size upon arrival of the target cell at time $0$ to be $Q^{\emph{CQ}}_{0}$ (or $Q^{\emph{OQ}}_{0}$), the CQ (or OQ) size upon its departure from the CQ at time $t$ to be $Q^{\emph{CQ}}_{t}$ (or $Q^{\emph{OQ}}_{t}$). Further assume the number of cells that arrive during time $0$ to $t$ is $Y_t$ (for both CQ and OQ), while the number of cells that are dropped in this period is $\Delta_t^{\emph{CQ}}$ for CQ (or $\Delta_t^{\emph{OQ}}$ for OQ). Finally, the number of cell departures from the CQ during time $0$ to $t$ is always $t$, because the target cell has not left yet and the work-conserving output always has something to serve. On the other hand, since the OQ switch is always more occupied than the CQ switch, it always has some cells to serve as well. Summing up, we get the following:
\begin{eqnarray}
Q_t^{\emph{CQ}}=Q_0^{\emph{CQ}}+Y_t-\Delta_t^{\emph{CQ}}-t,\\
Q_t^{\emph{OQ}}=Q_0^{\emph{OQ}}+Y_t-\Delta_t^{\emph{OQ}}-t.
\end{eqnarray}

The delay experienced by the target cell is $D^{\emph{CQ}}=Q_0^{\emph{CQ}}-\Delta_t^{\emph{CQ}}=t$ in the CQ switch, and  $D^{\emph{OQ}}=Q_0^{\emph{OQ}}-\Delta_t^{\emph{OQ}}=Q_t^{\emph{OQ}}-Y_t+t\ge Q_t^{\emph{CQ}}-Y_t+t=t$ in the OQ switch. Following the same argument as in 1), when the OCF policy is not used, as long as the scheduling policy is work conserving, the average delay experienced by each cell in the CQ switch is no higher than in the OQ switch. 

In conclusion, any work-conserving CQ switch always has an equal or better delay performance than the OQ switch if only accepted cells are taken into account, and does not suffer from the indefinite delay degradation problem due to output contentions as in many IQ switches. Simulation results in \cite{radonjic01,radonjic02} also support this conclusion.

We next investigate how the proposed buffering techniques will affect delay performance:

1) Load balancing, deflection routing and buffer pooling within a single daisy-chain only change the relative service order and cell drop rate, hence the delay performance is still bounded by that of an OQ switch;

2) Buffer pooling across multiple outputs may deteriorate the delay performance. If there is insufficient memory-read speedup, output contentions may cause indefinite delay and blocking. If there is sufficient speedup, the average delay may be larger but only because less cells may be dropped; upper bound may not hold if the drop rate is lower than that of \emph{OQ}.

\subsection {Support for Multicast}
In the following two sections, we discuss how multicast and QoS could be supported in the context of CQ switches.

Multicast has always been a concern in switch design over the decades. For an OQ switch, multicast suffers from the same factor-of-N speedup problem as unicast traffic, which makes it impractical for large switches. For an IQ switch, multicast makes the output contention problem even worse. When all traffic is unicast, a HOL cell may have to back off when any other HOL cell from another input target the same output, causing delays to itself and all cells behind it in the same queue. HOL blocking can be resolved by VOQs, but the delay performance is still affected and could be much worse than the OQ switch. For multicast traffic, scheduling becomes even more complicated \cite{multicast,iqmulticast}. 

Due to difficulties in supporting multicast in IQ and OQ switches, people have been looking at CQ switches in various contexts \cite{smoothscheduling,cqmulticast1}. Generally, CQ switches are especially suitable for multicast traffic due to its abundant input-output connections and distributed buffering modules:
1) Unlike in an IQ, OQ, or CIOQ switch, there is no need of extra memory speedup or fanout splitting to support multicast in CQ switches;
2) All admissible multicast traffic (i.e., no input over-subscription before replication, and no output over-subscription after replication \cite{multicast}) is naturally supported by CQ switches;
3) The only implementation concern is that we need to add a filtering and replication module at each crosspoint so that multicast cells can be selectively buffered.

Load balancing and deflection routing can be directly applied to multicast cases, and the proposed OCF and RR-based schemes can still maintain the correct order. For buffer pooling, multicast cells can still be replicated and directly sent to the corresponding buffer pools upon arrival at the input. Due to aggregation of crosspoints, multicast cells could be reused if they are destined to different outputs connected with the same buffer pool. Meanwhile, fanout splitting or no fanout splitting mechanisms may need to be applied within each buffer pool. 


\subsection{Quality of Service}
We next investigate how QoS could be supported in CQ switches. There has been abundant work on designing QoS-guaranteeing scheduling algorithms for the OQ switch. The reason why OQ is favored for QoS is that the OQ switch supports $100\%$ throughput without incuring extra delay, which is a property also shared by the basic CQ switch, but not the IQ switch. 

The basic CQ switch can work in the same way as the OQ switch, except that the queues of cells from different inputs are segregated, so QoS-guaranteeing algorithms suitable for OQ can also be applied to CQ. Load balancing and deflection routing shuffles the cells among crosspoint buffers associated with the same outputs. This may impede flow-level scheduling like LQF, but has no impact on cell-level scheduling like OCF.

The PCQ switch can be viewed as a compromise among IQ, OQ, CQ and SM switches, so does its QoS support. The PCQ switch generally cannot guarantee $100\%$ throughput, but may support both flow-level scheduling and cell-level scheduling. Additional buffer management rules can be adopted at each buffer pool, which specifies departure priorities and buffering partitions. For example, each buffer pool may decide which cell to serve by themselves when a service token is granted by some output port under the generalized LQF policy, and partial buffer sharing policies \cite{partitionsurvey} may help decide which cell to drop when the pool is full and provide another layer of service differentiation.

\section{Numerical Simulations}
\label{sec:simulation}
In this section, we use a C++ simulator to perform numerical simulations, and show the performance improvements through buffer sharing. Specifically, we compare the cell drop rates and critical buffer utilizations of the CCQ and PCQ switches against a basic LQF-based CQ switch and an OQ switch with the same total buffer space. The latter two systems are used as benchmarks in our comparison.

%
%
%
%



\subsection{Impact of Traffic Load on CCQ Switch}

First, we evaluate the effectiveness of load balancing and deflection routing under uniform bursty traffic. The destinations of incoming cells are evenly distributed among all $N$ output ports, i.e., $\lambda_{ij}=\frac{\mu}{N}$, $i,j=1,2,...,N$.

Since real Internet traffic is usually bursty and LRD, we focus on this kind of traffic. Specifically, we use the Markov Chain model in \cite{LRD} to generate LRD traffic with Hurst parameter $H=0.75$ and maximum length $L=1000$, i.e., each single burst of cells belonging to the same flow is restricted at most $1000$ time slots. Subsequently, we use this traffic-generating model, and adjust $H$, $L$, $\lambda_{ij}$ to control the traffic pattern.

We consider $32\times 32$ switches with crosspoint buffer size $B=40$ cells. The simulation lasts $T=10^7$ time-slots.

%

\begin{figure}[ht]
\begin{minipage}[t]{3.2 in}
\centering \subfigure{
\includegraphics[width=3.2 in]{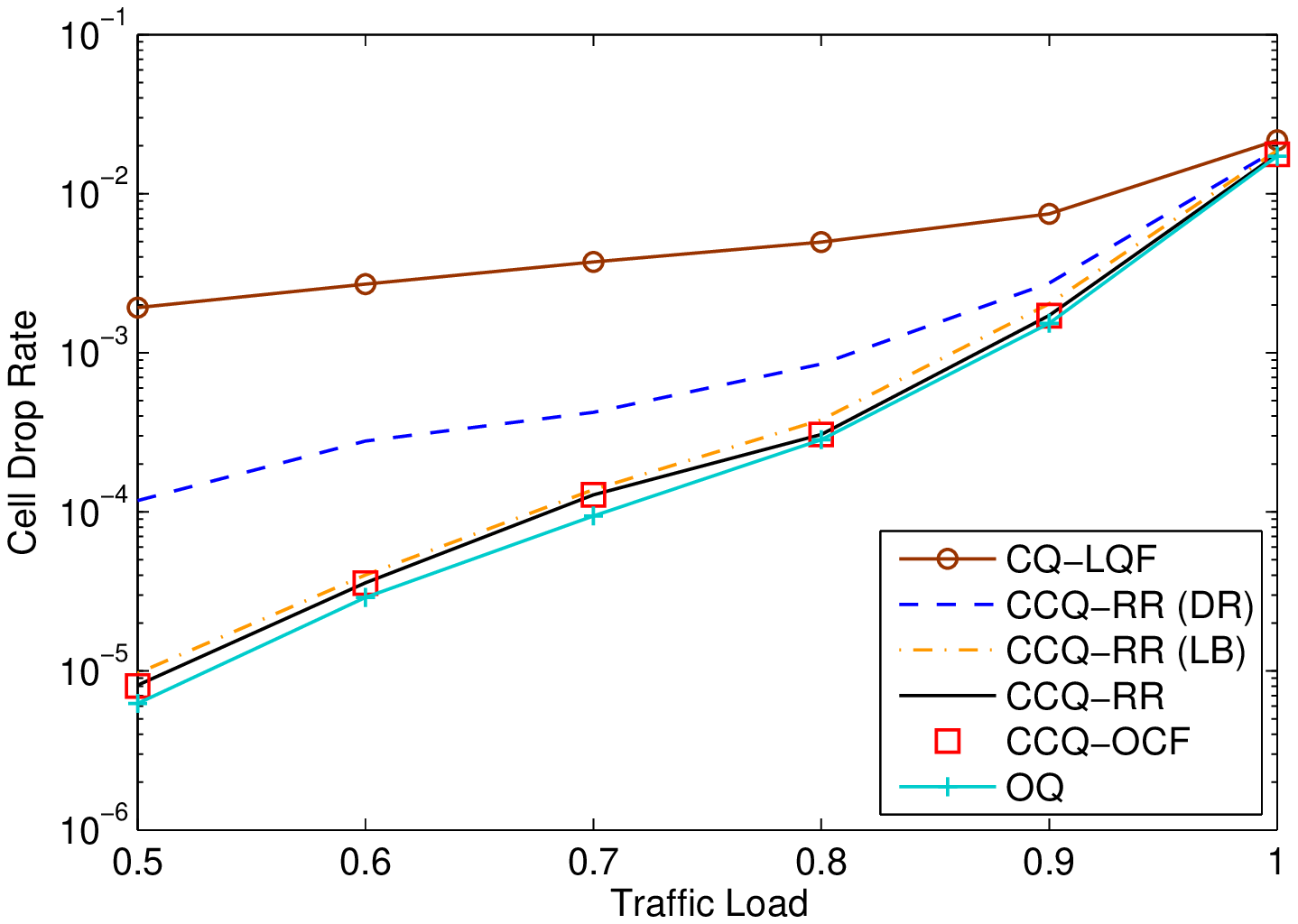}
\label{fig:drop_unif_bursty_1000}}
\end{minipage}
\begin{minipage}[t]{3.2 in}
\centering \subfigure{
\includegraphics[width=3.2 in]{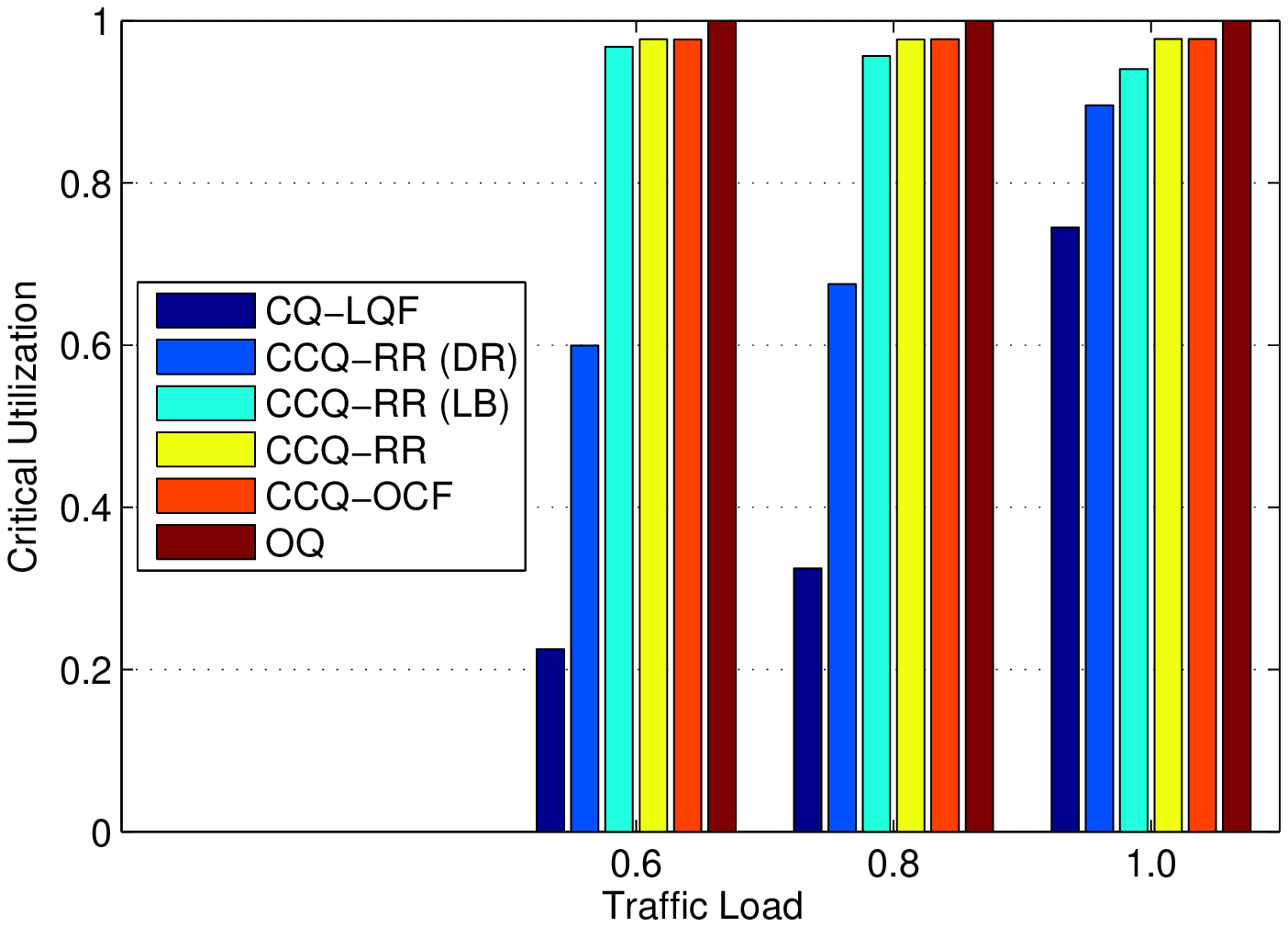}
\label{fig:utility_unif_bursty_1000}}
\end{minipage}
\caption{$32 \times 32$ CCQ switches with $B=40$ under uniform bursty traffic with $H=0.75$, $L=1000$, and $0.5\le \lambda \le 1.0$.}
\label{fig:unif}
\end{figure}

Fig. \ref{fig:drop_unif_bursty_1000} compares the cell drop rates of various schemes. The abbreviation ``\emph{CCQ-RR (LB)}'' (or ``\emph{CCQ-RR (DR)}'') stands for ``\emph{CCQ-RR} with load balancing (or deflection routing) only''. These two degenerate versions of \emph{CCQ-RR} are used to reveal the stand-alone effectiveness of load balancing and deflection routing.

Simulation results show that \emph{CCQ-OCF} and \emph{CCQ-RR} have the lowest cell drop rates, which are much better than that of \emph{CQ-LQF} and very close to that of \emph{OQ}, going down to about $10^{-5}$ when the traffic load $\mu=0.5$. Similar performances can also be achieved under higher traffic loads if larger buffers are implemented. 

Comparing \emph{CCQ-RR (LB)} with \emph{CCQ-RR (DR)}, we can find that deflection routing does not contribute as much as load balancing in this case. This is consistent with our analysis in Section \ref{sec:lb}, because load balancing transforms independent, bursty traffic into correlated, less-bursty traffic, while deflection routing only moves cell around regionally in short time-scales. However, one at this point cannot conclude that deflection routing is ineffective. In fact, the superiority of load balancing could largely be attributed to how we synthesize the LRD traffic. As mentioned before, our model generates separate bursts of cells that belong to different flows, which makes load balancing especially effective. On the other hand, in real Internet traffic, such bursts are often interleaved, showing Poisson characteristics in short time scales, and leaving more time for deflection routing to propagate. Besides, load balancing is a passive mechanism, while deflection routing is reactive and complementary.

We also compare the buffer utilizations of different schemes in Fig. \ref{fig:utility_unif_bursty_1000}. Here we can see that the critical utilization of \emph{CQ-LQF} is fair when the traffic load is high (about $70\%$ when $\mu=1.0$), but drops quickly as the traffic load is reduced (only $20\%$ when $\mu=0.6$). To understand this, we must realize that a lower traffic load does not necessarily lead to less burstiness according to our model, since the Hurst parameter does not change at all. Ironically, when the traffic load is lower, the incoming traffic at different crosspoints can be even more unbalanced in a short time-scale. This kind of low buffer utilization leads to a larger performance degradation when the traffic load is low (as compared with \emph{OQ}). This is also consistent with our analysis in Section \ref{sec:largebuffer}. By contrast, \emph{CCQ-OCF} and \emph{CCQ-RR} are not affected by the change of traffic load, showing robustness against various traffic loads.

Comparing Fig. \ref{fig:drop_unif_bursty_1000} and Fig. \ref{fig:utility_unif_bursty_1000}, we can see a clear trend that the cell drop rate is negatively correlated with the critical buffer utilization. The critical buffer utilizations of \emph{CCQ-OCF} and \emph{CCQ-RR} are close to $100\%$, which is only achievable by the OQ switch. Thus the significant performance improvements of the proposed schemes can be attributed to their efficient buffer sharing mechanisms.

\subsection{Impact of Non-uniformity on CCQ Switch}

In addition to the uniform bursty traffic, we also test the proposed buffer sharing and scheduling techniques under non-uniform traffic. In this case, we adopt a hot-spot traffic model in which $\lambda_{ii}=a\mu$, and $\lambda_{ij}=\frac{(1-a)\mu}{N-1}$ for $i \neq j$,
where $a$ is the hot-spot factor.
We still focus on $32\times 32$ CQ switches with buffer size $B=40$ cells. The incoming traffic is LRD with $H=0.75$, $L=1000$, and $a=0.5$.

%

\begin{figure}[ht]
\begin{minipage}[t]{3.2 in}
\centering \subfigure{
\includegraphics[width=3.2 in]{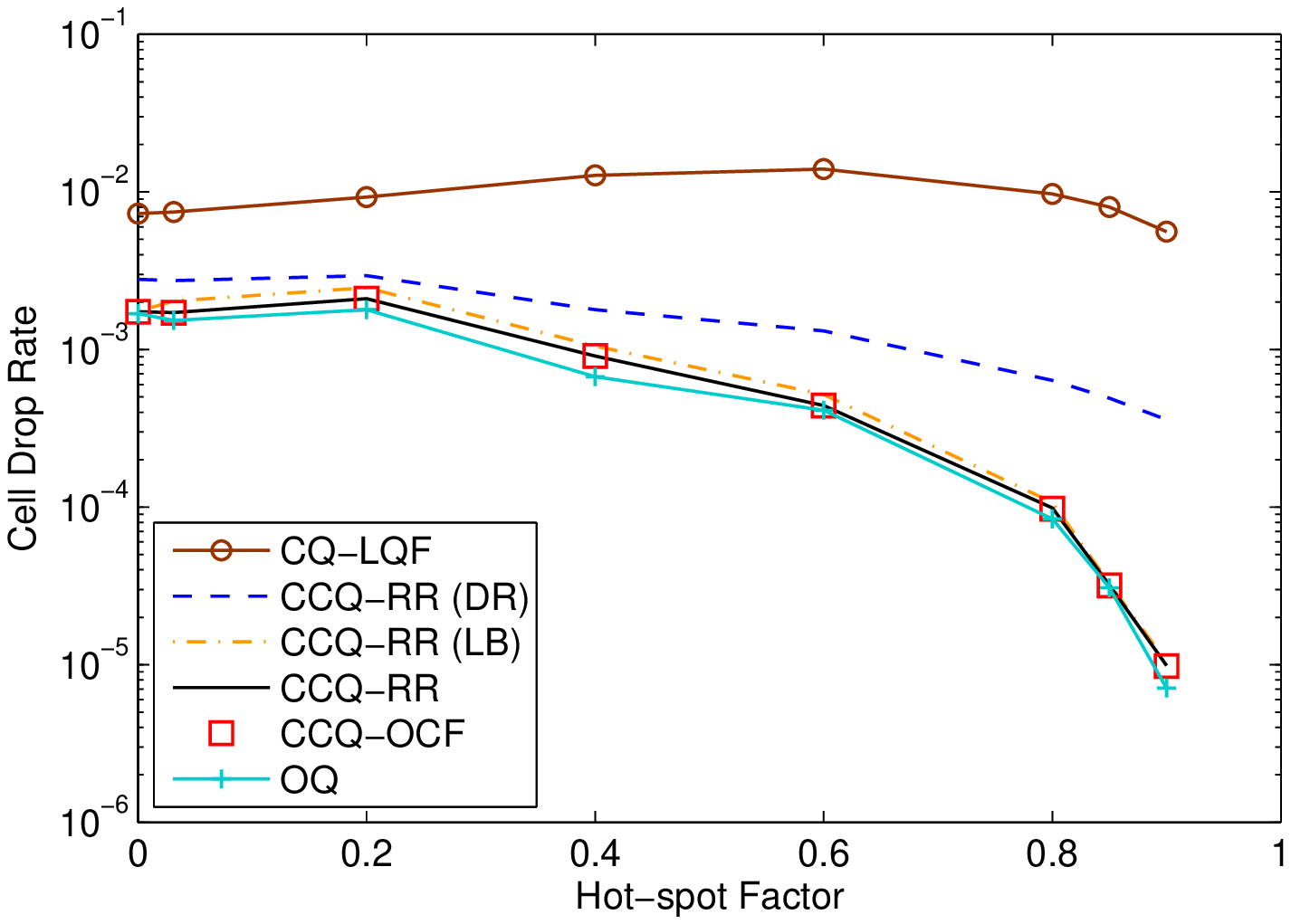}
\label{fig:drop_nonunif_bursty}}
\end{minipage}
\begin{minipage}[t]{3.2 in}
\centering \subfigure{
\includegraphics[width=3.2 in]{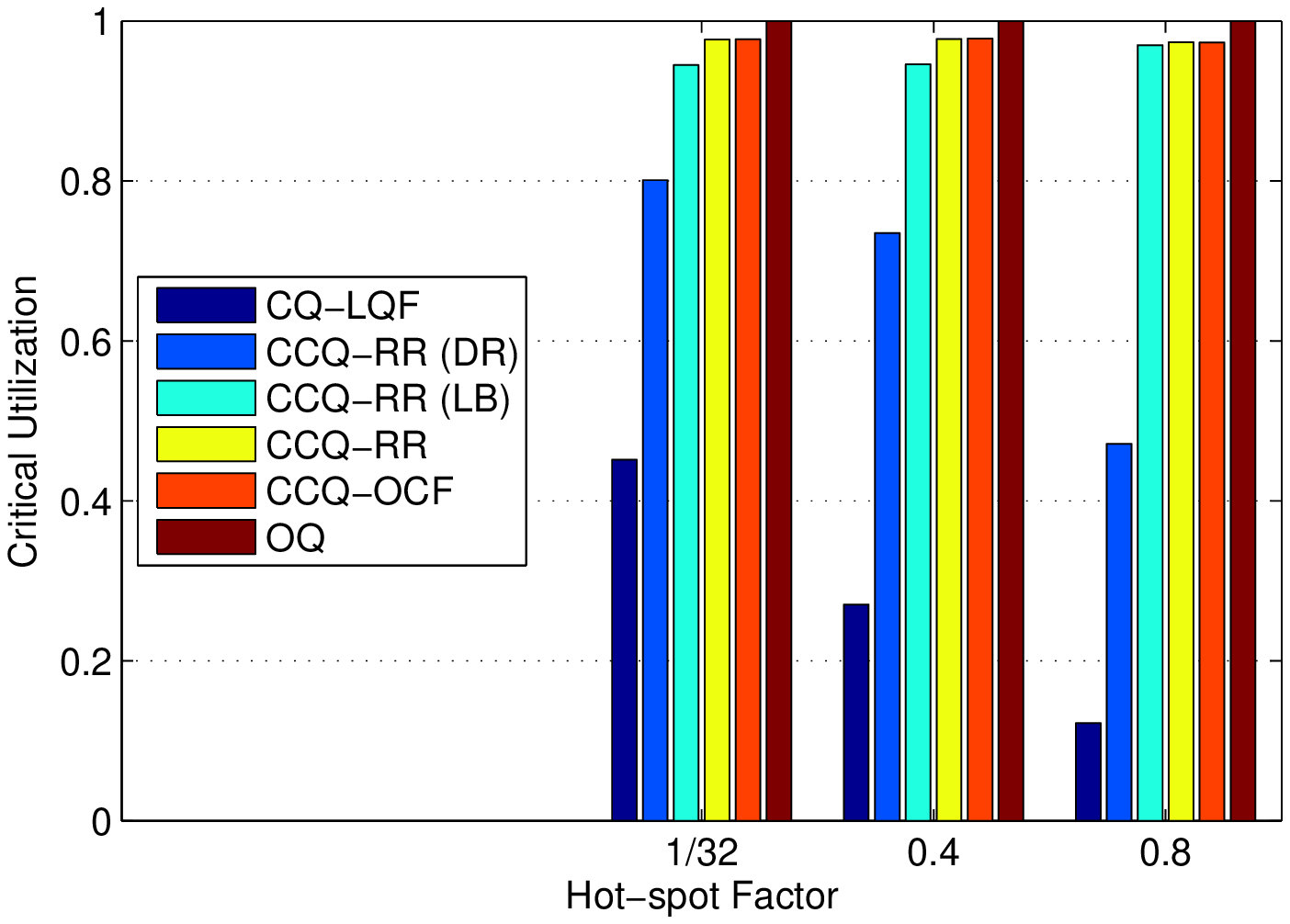}
\label{fig:utility_nonunif_bursty}}
\end{minipage}
\caption{$32 \times 32$ CCQ switches with $B=40$ under non-uniform bursty traffic with $\lambda=0.9$, $H=0.75$, $L=1000$, and $0\le a \le 0.9$.}
\label{fig:nonunif}
\end{figure}

The cell drop rates and critical buffer utilizations of the proposed schemes under hot-spot LRD traffic are illustrated in Fig. \ref{fig:nonunif}. From these two figures, we find that non-uniformity does not significantly hurt the performances. Instead, the cell drop rate may drop dramatically when the hot-spot factor is large. In fact, when $a=1$, there would be a perfect one-to-one matching between all input-output pairs, and there would be no cell drops for any work-conserving scheduling policy. However, different switch configurations and scheduling algorithms behave differently as $a$ grows larger. The cell drop rate of \emph{CQ-LQF} slightly increases when $\frac{1}{32}\le a \le 0.6$ and then drops slowly after that, reaching $5\times 10^{-3}$ at $a=0.9$. Meanwhile, its critical buffer utilization consistently decreases as the traffic becomes more non-uniform. This can be attributed to the fact that LQF is rate-unaware and cannot perfectly identify the queue that is most likely to overflow under non-uniform traffic. Also, the buffer utilizations could be more unbalanced in this case because only one crosspoint buffer is frequently used while all others are always under-utilized. By contrast, \emph{CCQ-RR} and \emph{CCQ-OCF} derive as much benefit from the non-uniformity as \emph{OQ} does, reaching $10^{-5}$ cell drop rate when $a=0.9$, with their critical buffer utilization are always close to 1. These results show that the proactive load balancing and reactive deflection routing are capable of combating non-uniformity, and perform relatively better under non-uniform traffic. We also notice that deflection routing by itself cannot fully handle unbalanced traffic, so load balancing is especially necessary for such traffic. Therefore, the conclusion in Section \ref{sec:framework} that load balancing is the best strategy to combat non-uniformity is validated.


\subsection{Impact of Burstiness on CCQ switch}

The impact of burstiness on the performance of CCQ switches is also investigated. Here we fix the maximum length to $L=1000$, then vary the Hurst parameter in the range $0.6\le H\le 0.9$.

%

\begin{figure}[ht]
\begin{minipage}[t]{3.2 in}
\centering \subfigure{
\includegraphics[width=3.2 in]{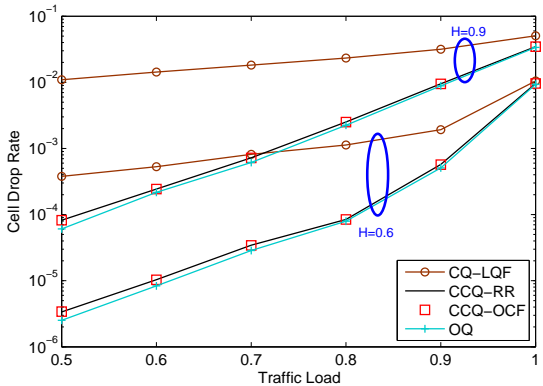}
\label{fig:drop_H}}
\end{minipage}
\begin{minipage}[t]{3.2 in}
\centering \subfigure{
\includegraphics[width=3.2 in]{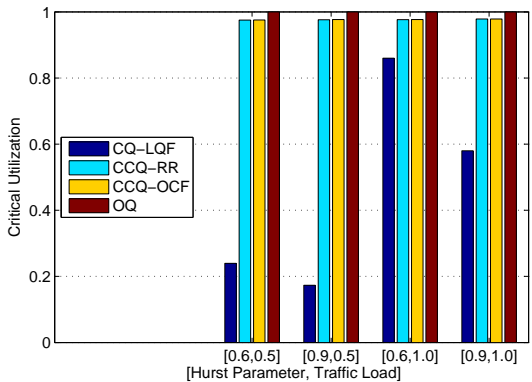}
\label{fig:utility_H}}
\end{minipage}
\caption{$32 \times 32$ CCQ switches with $B=40$ under uniform bursty traffic with $0.6\le H\le 0.9$, $L=1000$, and $0.5\le \lambda \le 1.0$.}
\label{fig:H}
\end{figure}

Simulation results in Fig. \ref{fig:H} show that \emph{CQ-LQF} performs worse when the traffic is more bursty but of lower load. On the other hand, the proposed \emph{CCQ-OCF} and \emph{CCQ-RR} schemes are not affected much, demonstrating their robustness against different burstiness levels. The underlying reason is that the small crosspoint buffers become less capable to sustain the traffic fluctuations as the incoming cells become more bursty and intermittent, and depend more on load balancing and deflection routing to smooth the traffic. Meanwhile, the LQF policy is not aware of the burstiness, and thus cannot always identify the crosspoint that is most likely to overflow. Also, LQF decreases the length(s) of the longest queue(s) only, unlike load balancing and deflection routing that extend their scope to shorter queues throughout the daisy chains as well. As a result, the proposed schemes gain relatively larger advantages under highly bursty traffic.

\subsection{Impact of Buffer Size on CCQ Switch}

Till now we have been using the same switch configurations, and examine their performances under various traffic patterns. In the following, we fix the incoming traffic pattern instead, and study the impact of buffer size and switch size on these switches.

\begin{figure}[ht]
\begin{minipage}[t]{3.2 in}
\centering \subfigure{
\includegraphics[width=3.2 in]{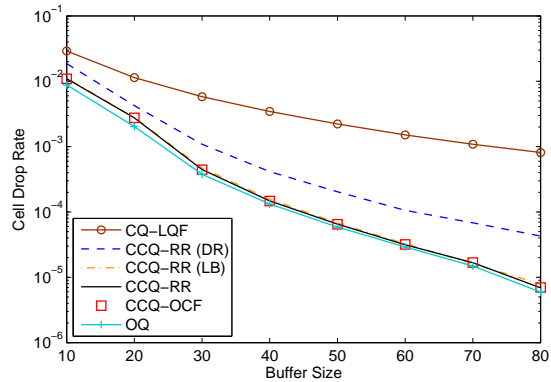}
\label{fig:drop_B}}
\end{minipage}
\begin{minipage}[t]{3.2 in}
\centering \subfigure{
\includegraphics[width=3.2 in]{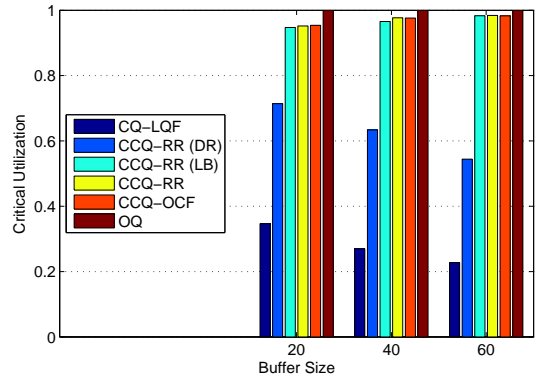}
\label{fig:utility_B}}
\end{minipage}
\caption{$32 \times 32$ CCQ switches with $10\le B\le 80$ under uniform bursty traffic with $H=0.75$, $L=1000$, and $\lambda =0.7$.}
\label{fig:B}
\end{figure}

The cell drop rates and critical buffer utilizations under various buffer sizes are plotted in Fig. \ref{fig:B}. It is evident from Fig. \ref{fig:drop_B} that the logarithmic decay rate of the cell drop rate with respect to the buffer size is always sub-linear, irrespective of the switch configurations and scheduling policies. This is the result of long-range dependence, as predicted in Section \ref{sec:inefficiency}. Even though all switches suffer from long-range dependence, the proposed CCQ switches have much deeper curves than \emph{CQ-LQF}, and is always close to that of \emph{OQ}. 

The same result can also be drawn from Fig. \ref{fig:utility_B}. The inefficiency of the LQF policy becomes more evident when the buffer size becomes larger. This may look inconsistent from the asymptotic analysis derived under uniform Bernoulli i.i.d. traffic in Section \ref{sec:inefficiency}, which states that the critical buffer utilization should tend to a constant when the buffer size grows to infinity. However, note that the effect of increasing buffer size is sublinear for LRD traffic, which is in accordance with our prior expectations.

\subsection{Impact of Switch Size on CCQ Switch}

The impact of buffer size has just been studied. What if the switch becomes larger, i.e., with more input and output ports? Here we investigate the impact of large $N$ on different switch configurations by fixing the total amount of buffer size per output, and consider a larger $128\times 128$ CQ switch with a smaller crosspoint buffer size of $10$.

%

\begin{figure}[ht]
\begin{minipage}[t]{3.2 in}
\centering \subfigure{
\includegraphics[width=3.2 in]{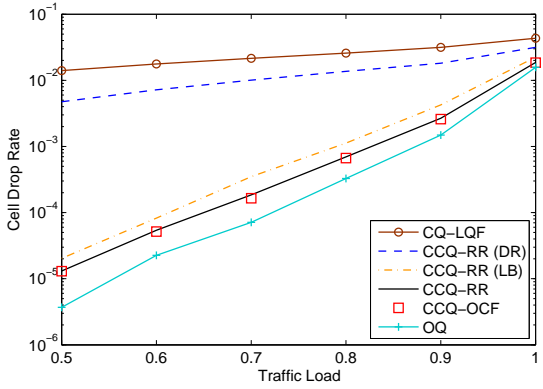}
\label{fig:drop_N}}
\end{minipage}
\begin{minipage}[t]{3.2 in}
\centering \subfigure{
\includegraphics[width=3.2 in]{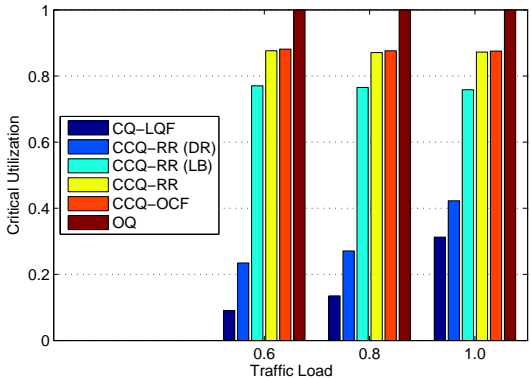}
\label{fig:utility_N}}
\end{minipage}
\caption{$128 \times 128$ CCQ switches with $B=10$ under uniform bursty traffic with $H=0.75$, $L=1000$, and $0.5\le \lambda \le 1.0$.}
\label{fig:N}
\end{figure}

From Fig. \ref{fig:N}, we can see that the legacy \emph{CQ-LQF} method suffers from a higher cell drop rate due to a smaller crosspoint buffer size. \emph{CCQ-OCF} and \emph{CCQ-RR} gain a larger advantage over \emph{CQ-LQF} in this case, but are inferior to \emph{OQ} due to increased difficulties in buffer sharing along longer daisy chains of smaller crosspoint buffers. Notwithstanding this issue, we may still claim that the proposed schemes are more useful for large switches with small crosspoint buffers. We also notice that deflection routing becomes much less effective when $N$ grows larger, because its buffer sharing effect is short-range and requires more time to propagate than load balancing.

A larger switch size of $N=128$ needs additional buffer space to achieve the same satisfactory cell drop rates as before. For \emph{CQ-LQF}, the total buffer space required to achieve similar performances may scale as $\Theta(N^2)$, since each crosspoint buffer must at least tolerate a single burst, whose length does not shrink much as $N$ increases. By contrast, for \emph{CCQ-OCF}, \emph{CCQ-RR} and \emph{OQ}, the total buffer space required to achieve similar performances does not scale so poorly. Even though the switch size is $4$ times larger than before, the aggregated buffer size for each output does not change at all, i.e., $N\times B \equiv 1280$ cells, and the total buffer space of all outputs scales as $\Theta(N)$.

For an OQ switch, this is easy to understand. Since the traffic load at each output always equals $\mu$, and if a Poisson arrival process is assumed, the output queue length distributions are always the same, irrespective of $N$. The LRD arrival process is certainly different, but as long as the burst length is not too large compared with the output buffer size, the performance of \emph{OQ} stays approximately the same. \emph{CCQ-OCF} and \emph{CCQ-RR} also share the segregated crosspoint buffers efficiently. That is why the total amount of buffers in each daisy chain stays almost the same for a given traffic level and loss performance, irrespective of the change in switch size.

\subsection{Impact of Pooling Pattern on PCQ Switch}

In this part, we investigate how buffer pooling affects the cell drop rates and buffer utilizations. $32\times 32$ PCQ switches with the same pooling size $w\times r = 8$ and full speedups but different pooling patterns are compared.

\begin{figure}[ht]
\begin{minipage}[t]{3.2 in}
\centering \subfigure{
\includegraphics[width=3.2 in]{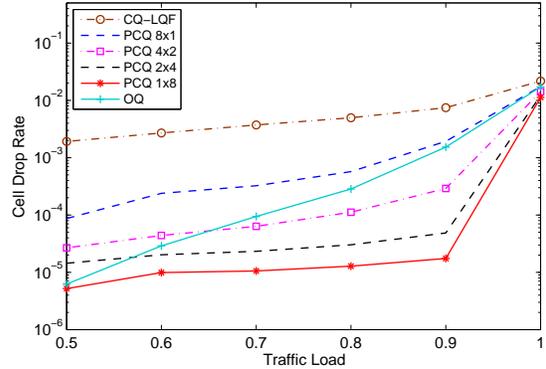}
\label{fig:drop_PCQ}}
\end{minipage}
\begin{minipage}[t]{3.2 in}
\centering \subfigure{
\includegraphics[width=3.2 in]{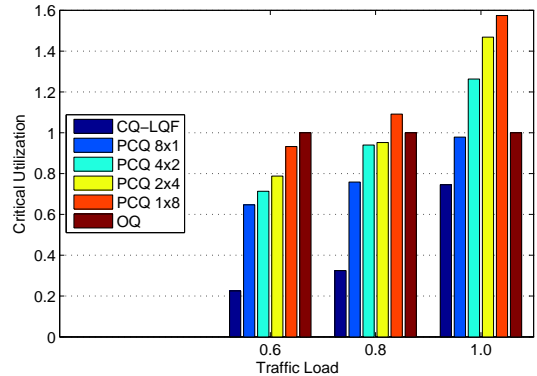}
\label{fig:utility_PCQ}}
\end{minipage}
\caption{$32 \times 32$ PCQ switches with $B=40$ under uniform bursty traffic with $H=0.75$, $L=1000$, and $0.5\le \lambda \le 1.0$.}
\label{fig:PP}
\end{figure}

In Fig. \ref{fig:drop_PCQ}, it can be found that buffer pooling is more efficient with a larger $r$ and a smaller $w$, which supports our conclusion in Section \ref{sec:wxr} that pooling crosspoints with shared inputs but different outputs is more effective in reducing the cell drop rate. Notice that \emph{PCQ-GLQF} performs better when the traffic load is high, and the gradient of these curves are much flatter than that of \emph{OQ}. This can be attributed to the fact that the dominant cause of cell drops under high traffic load is simultaneous cell arrival from many inputs, whereas the length of a single burst plays a more important role under low traffic load. This is consistent with our analysis in Section \ref{sec:largebuffer}.

Similar insights can also be drawn from Fig. \ref{fig:utility_PCQ}. One interesting phenomenon is that \emph{PCQ-GLQF} actually pushes the limit of buffer sharing across different outputs, and may sometimes break through the limit of $100\%$ critical utilization, because crosspoints associated with a busy output may temporarily borrow some buffer space from its neighbors that are in the same pool but associated with a less congested output. In the extreme case, under $1\times m$ buffer pooling, a single crosspoint may borrow up to $m-1$ times of its normal buffer size $B$ from its pooling neighbors, and thus a single output may take up to $mNB$ cells of buffer space before experiencing an overflow. This explains why \emph{PCQ-GLQF} may outperform load balancing, deflection routing and even \emph{OQ} in some scenarios.

\subsection{Impact of Memory Speedup on PCQ Switch}

In the previous discussion, full speedup is assumed for each pooling pattern, which could be inefficient and unnecessary. Here we examine the performance of \emph{PCQ-GLQF-MWM} schemes when memory speedup is restricted.
\begin{figure}[ht]
\begin{minipage}[t]{3.2 in}
\centering \subfigure{
\includegraphics[width=3.2 in]{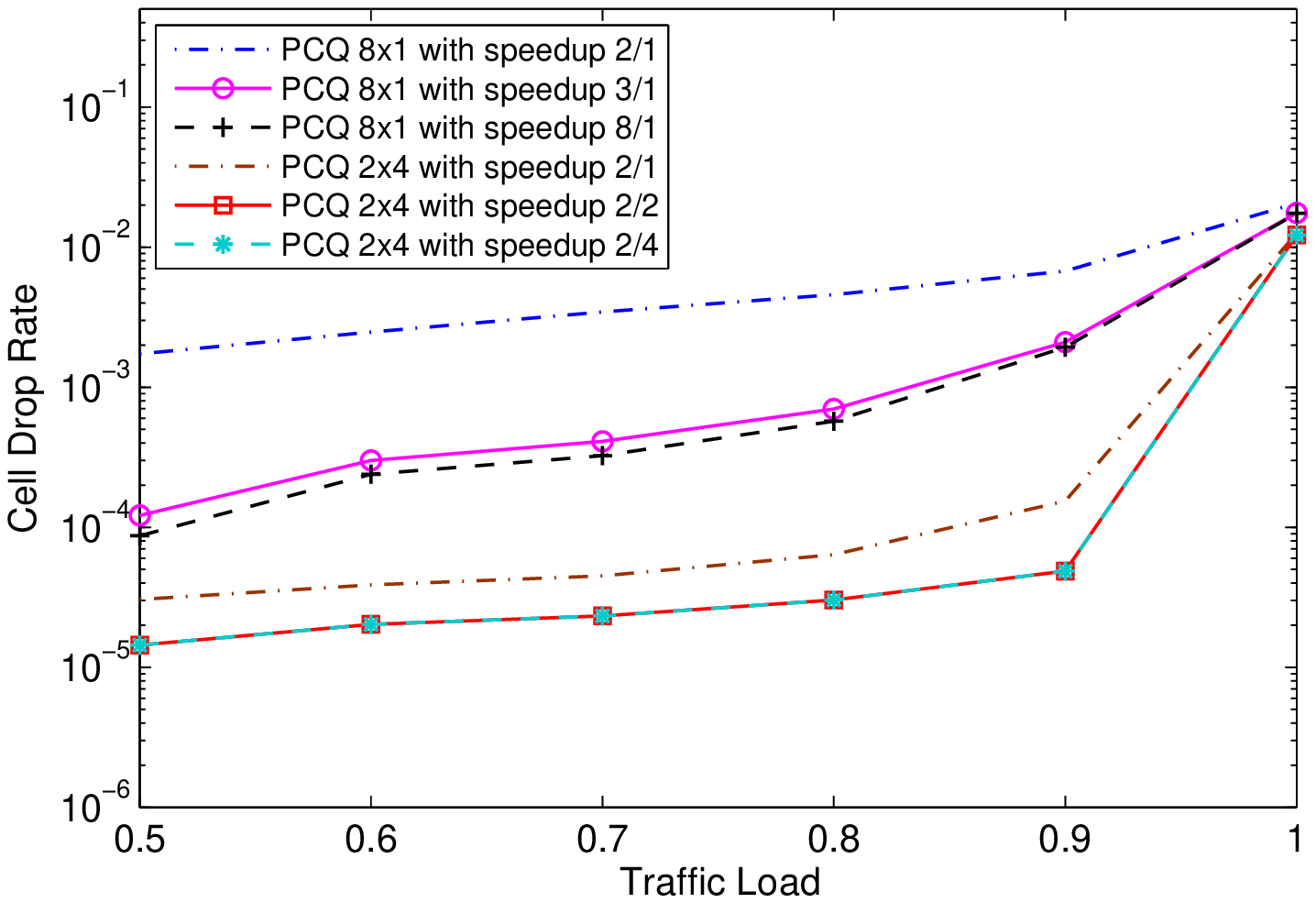}
\label{fig:drop_PCQ_speedup}}
\end{minipage}
\begin{minipage}[t]{3.2 in}
\centering \subfigure{
\includegraphics[width=3.2 in]{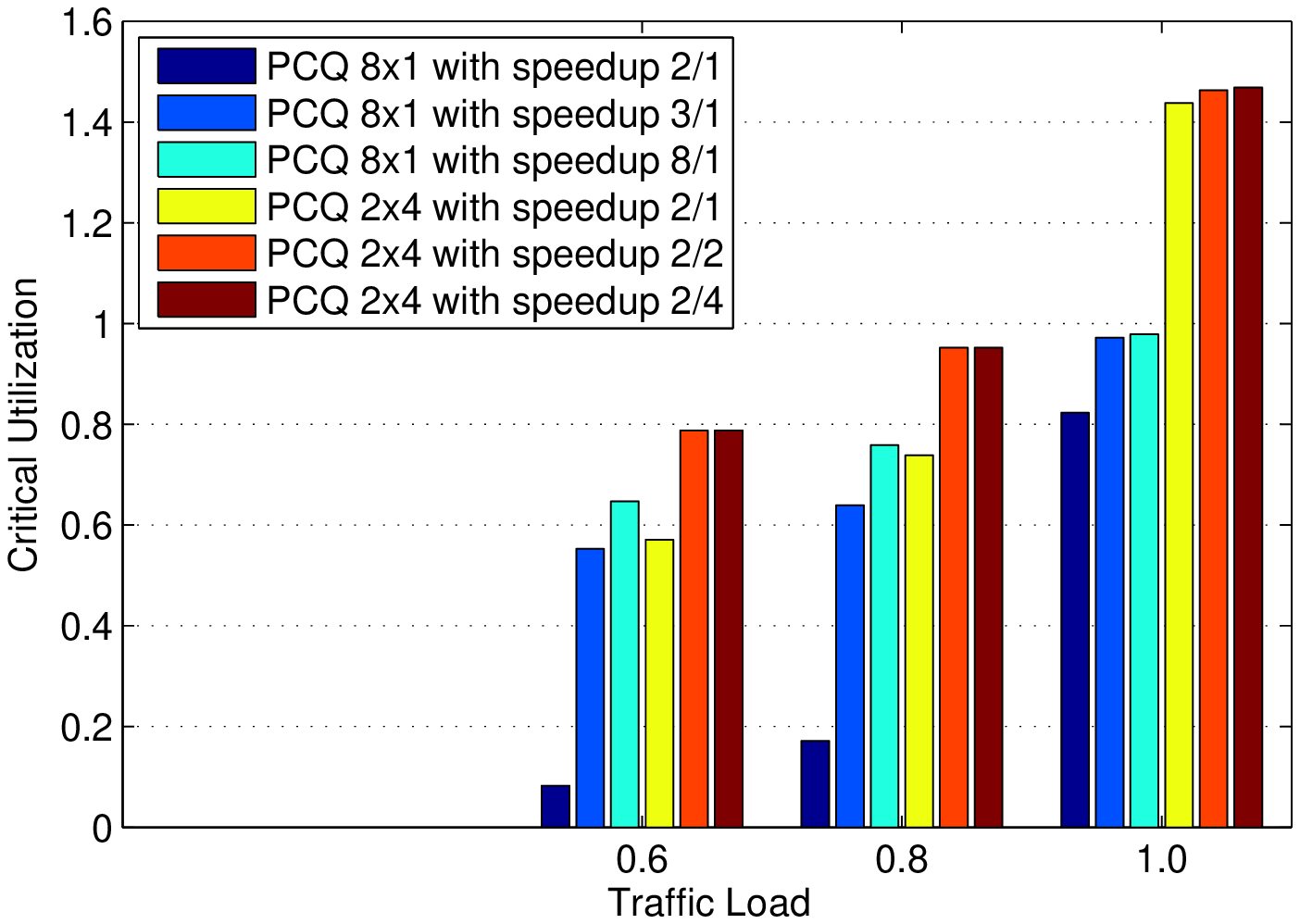}
\label{fig:utility_PCQ_speedup}}
\end{minipage}
\caption{$32 \times 32$ PCQ switches with $B=40$ under uniform bursty traffic with $H=0.75$, $L=1000$, and $0.5\le \lambda \le 1.0$.}
\label{fig:speedup}
\end{figure}

Fig. \ref{fig:speedup} shows how insufficient memory-write/read speedup impacts the performance of PCQ switches. For $8\times 1$ pooling, a low memory-write speedup of $s_w=3$ results in a similar performance as a full speedup of $s_w=w=8$, which is due to the low probability of simultaneous cell arrivals from multiple inputs. For $2\times 4$ buffer pooling, the contention resolution mechanism diminishes the need for a memory-read speedup. In addition, a memory-read speedup of $s_r>s_w$ has almost no effect in improving the performance, as indicated by the overlapping curves of $s_r=2$ and $s_r=r=4$. All these observations are consistent with our analysis in Section \ref{sec:contention}.

\subsection{Real Internet Traces}

Finally, we test the proposed schemes using real Internet traces. In the simulation, a different CAIDA OC-192 ($10Gbps$) trace \cite{ipv6day} is fed into each input port of the CQ switch. The incoming packets are hashed according to a fixed look-up table, so that all outputs receive approximately the same traffic load. Variable-length IP packets are fragmented into fixed-length cells of $64 byte$ each, which is a typical value used in Internet core switches. 

%

\begin{figure}[ht]
\begin{minipage}[t]{3.2 in}
\centering \subfigure{
\includegraphics[width=3.2 in]{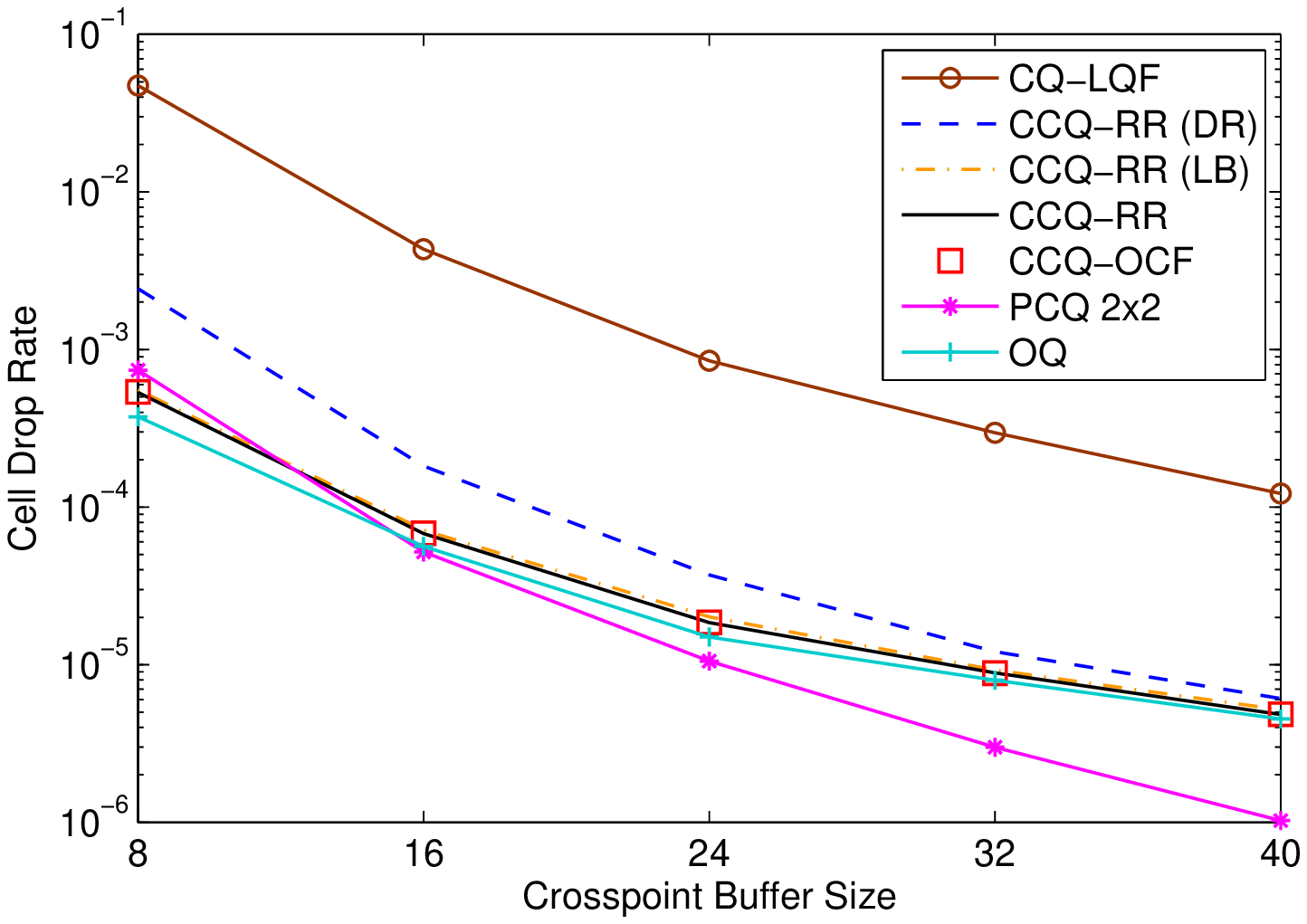}
\label{fig:drop_caida}}
\end{minipage}
\begin{minipage}[t]{3.2 in}
\centering \subfigure{
\includegraphics[width=3.2 in]{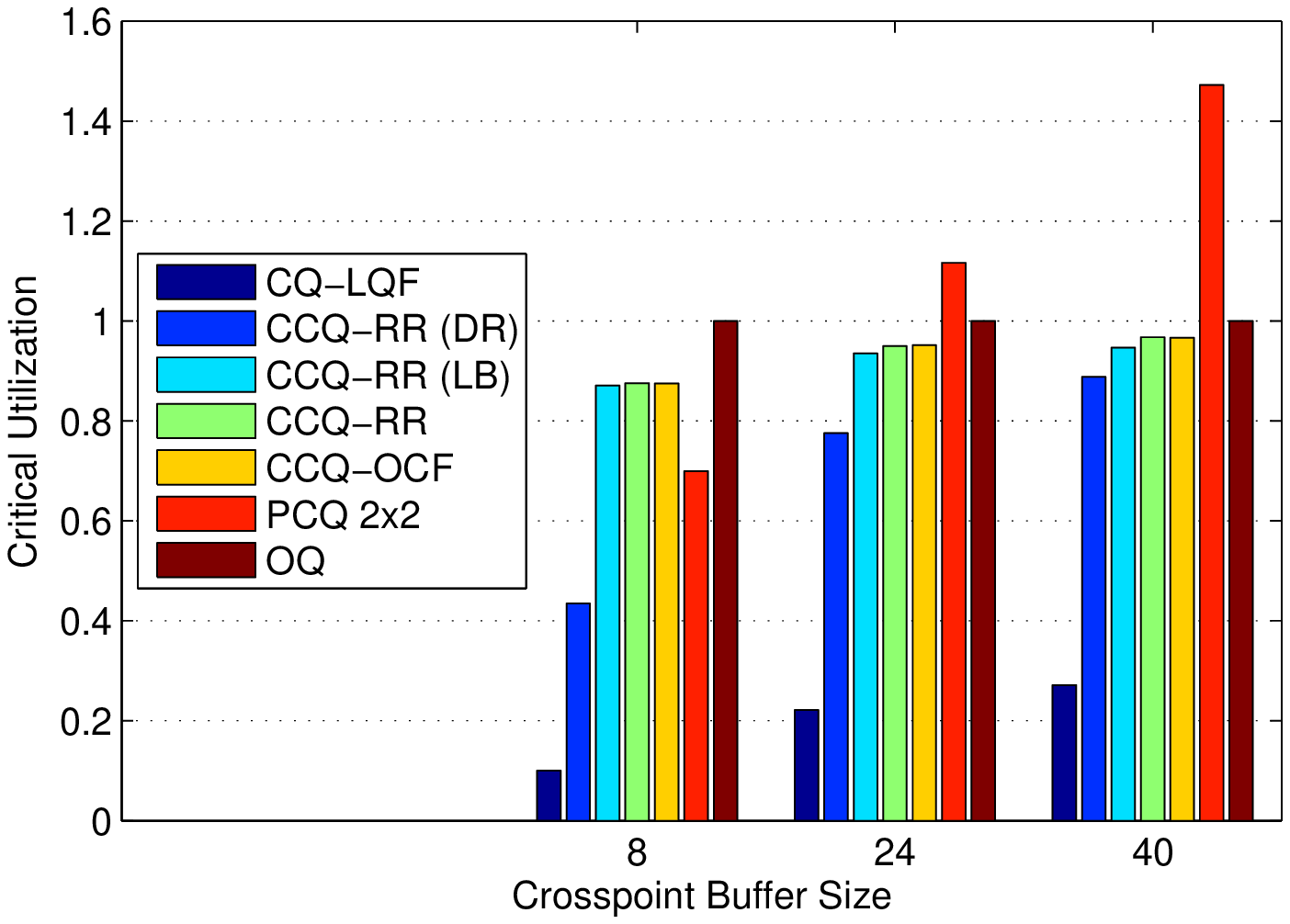}
\label{fig:utility_caida}}
\end{minipage}
\caption{$32 \times 32$ switches with $8\le B\le 48$ under real Internet traces with $\lambda\approx 0.45$ and $H \approx 0.75$.}
\label{fig:caida}
\end{figure}

First we consider $32\times 32$ CQ switches, and use the original traces from CAIDA with an average traffic load of $\mu \approx 0.45$ and a measured Hurst parameter of $H\approx 0.75$. The simulation period is $T=10^7$ time slots. Examination of the packet headers reveals that over $50,000$ flows with different source/destination IP addresses are multiplexed into each link during the simulation period. As displayed in Fig. \ref{fig:caida}, \emph{CCQ-OCF} and \emph{CCQ-RR} ensure very low cell drop rates, about $10$ to $100$ times lower than the basic LQF-based CQ switch, and close to the OQ switch with the same total buffer space. Also note that deflection routing contributes more as the crosspoint buffer size grows larger. Furthermore, \emph{PCQ-GLQF} achieves even better performances than \emph{OQ} with just a small pooling size of $2\times 2$ and memory speedup $s_w/s_r=2/1$. To support an average cell drop rate of $10^{-5}$, only about $32\times 32\times 40\times64 byte=2.5 Mbyte$ total buffer space is needed.


%

\begin{figure}[ht]
\begin{minipage}[t]{3.2 in}
\centering \subfigure{
\includegraphics[width=3.2 in]{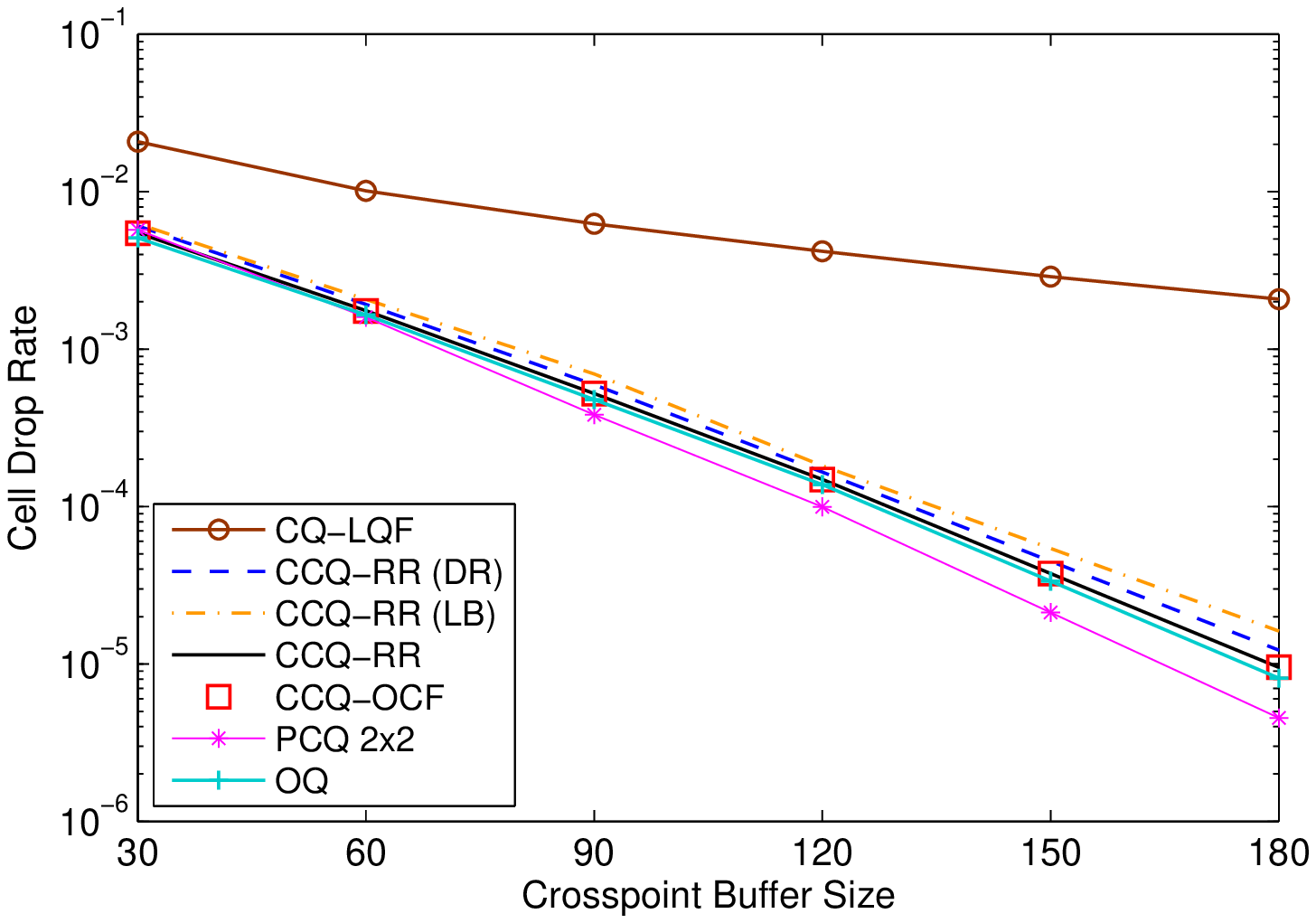}
\label{fig:drop_caida09}}
\end{minipage}
\begin{minipage}[t]{3.2 in}
\centering \subfigure{
\includegraphics[width=3.2 in]{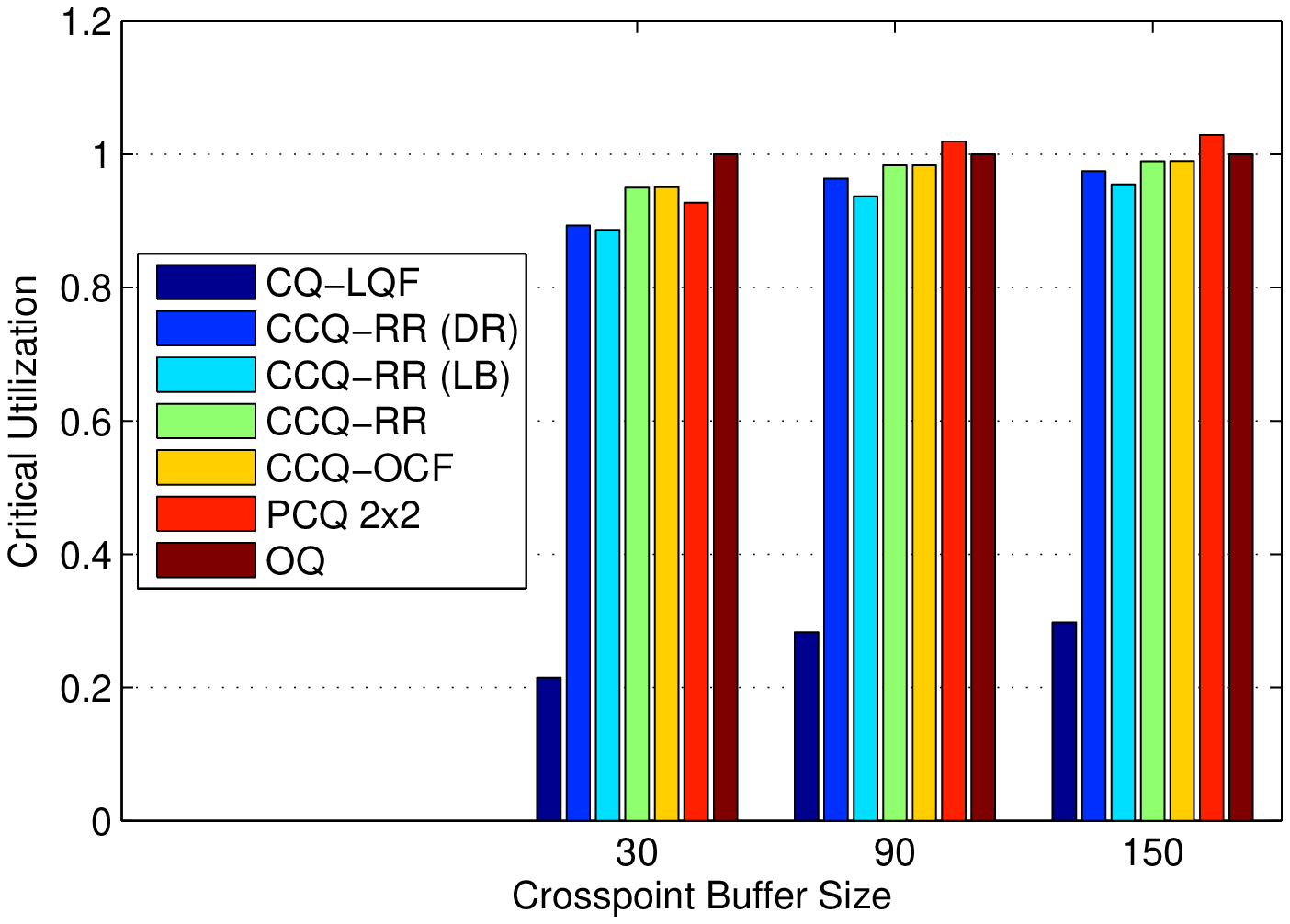}
\label{fig:utility_caida09}}
\end{minipage}
\caption{$128 \times 128$ switches with $30\le B\le 180$ under real Internet traces with $\lambda=0.9$.}
\label{fig:128}
\end{figure}

We then consider a larger $128\times 128$ CQ switch. We use the same Internet traces, but reduce the core switching speed and place throttles right before the input ports so that the system effectively works at a higher traffic load of $\mu=0.9$. The cell drop rates and buffer utilizations are shown in Fig. \ref{fig:128}. In this case, a much larger memory space, $128\times 128\times 180 \times 64 byte=180 Mbyte$, is required to achieve the same cell drop rate of $10^{-5}$, but it is still feasible using the state-of-art ASIC technologies \cite{asic01,asic02,asic03}. In fact, by pushing the on-chip bytes to the limit of $455 Mbyte$, we may extrapolate the curves in Fig. \ref{fig:drop_caida09} and conjecture that an even lower drop rate of $10^{-8}$ can be achieved. The relative performance gains of the proposed schemes over \emph{CQ-LQF} are even higher in this case. Also note that the deflection routing mechanism in \emph{CCQ-RR} works better than load balancing in this case.

\section{Conclusion}
\label{sec:conclusion}
In this paper, we address the crucial buffering constraints in a single-chip CQ switch. At the cost of some modest hardware modifications and memory speedup, we make it possible for the segregated buffers at different crosspoints to be dynamically shared along daisy chains which effectively mimics an OQ switch, or within buffer pools that enable buffer sharing across multiple inputs and outputs. We also propose novel scheduling schemes that can maintain the correct packet ordering with low complexity and resolve contentions with low speedup, which are also important in designing packet-switched networks. Exploiting the benefits of load balancing, deflection routing, and buffer pooling, we significantly improve the buffer utilizations by up to $10$ times and reduce the packet drop rates by one to three orders of magnitude, especially for large switches with small crosspoint buffers under bursty and non-uniform traffic. Extensive simulations have been performed to demonstrate that the memory sizes available using current ASIC technology is sufficient to deliver a satisfactory performance with a single-chip CQ architecture.

%

\ifCLASSOPTIONcaptionsoff
  \newpage
\fi


%




\end{document}